\documentclass[10pt]{article}
\usepackage{amsfonts}
\usepackage{amssymb}
\usepackage{amsthm}

\usepackage[pdftex]{graphicx}

\newcommand{\sss}{\setcounter{equation}{0}}
\newtheorem{theorem}{THEOREM}[section]

\newtheorem{lemma}[theorem]{LEMMA}

\newtheorem{corollary}[theorem]{COROLLARY}

\newtheorem{remark}[theorem]{REMARK}

\newtheorem{definition}[theorem]{DEFINITION}

\newtheorem{assumption}[theorem]{ASSUMPTION}

\newcommand{\ap }{\mathbf A}
\newcommand{\ere}{ {\mathbb R}}
\newcommand{\ZETA}{{\mathbb Z}}
\newcommand{\ese}{{\mathbb S}}

\def\p2{\mathcal A_{\Phi,2\pi}(B)}
\def\0p2{\mathcal A_{\Phi,2\pi}(0)}
\def\sp2{\mathcal A_{\Phi,2\pi,\hbox{\rm SR}}(B)}
\def\beq{\begin{equation}}
\def\ene{\end{equation}}
\def \ds {\displaystyle}
\newcommand{\bull}{\hfill $\Box$}
\def\qed{\ifhmode\unskip\nobreak\fi\ifmmode\ifinner
\else\hskip5pt\fi\fi\hbox{\hskip5pt\vrule width4pt height6pt
depth1.5pt\hskip1pt}}
\def\v{\mathbf v}

\def\hv{\hat{\mathbf v}}

\def\hw{\hat{\mathbf w}}
\def\curl{\, \hbox{ \rm curl}\,}
\def\mo{\mathbf p}
\def\ta{\tilde{A}}

\def\tvf{\tilde{\varphi}}

\def\div{\,\hbox{\rm div}\,}
\def\xin{x_{\rm in}}
\def\xout{ x_{\rm out}}
\def\+out{x^{\rm out}}
\begin{document}
\baselineskip=20 pt
\parskip 6 pt

\title{Aharonov-Bohm Effect and High-Velocity Estimates of
Solutions to the Schr\"odinger Equation
\thanks{ PACS Classification (2008): 03.65Nk, 03.65.Ca, 03.65.Db, 03.65.Ta.  Mathematics Subject Classification(2000): 81U40, 35P25,
35Q40, 35R30.}
\thanks{ Research partially supported by
 CONACYT under Project Problemas Matem\'aticos de la F\'{i}sica Cu\'antica.}}
\author{ Miguel Ballesteros \thanks{ Electronic Mail: ballesteros.miguel.math@gmail.com}\\
 Johannes Gutenberg-Universit\"at.\\
Institut f\"ur Mathematik. Staudingerweg 9
55099 Mainz. Germany
\and Ricardo Weder\thanks {Fellow, Sistema Nacional de Investigadores.
Electronic mail: weder@servidor.unam.mx  }\\
Departamento de M\'etodos
 Matem\'aticos  y Num\'ericos.\\
 Instituto de Investigaciones en Matem\'aticas Aplicadas y en
 Sistemas. \\
 Universidad Nacional Aut\'onoma de M\'exico.\\
  Apartado Postal 20-726,
M\'exico DF 01000, M\'exico.}

\date{}
\maketitle

  \centerline{To Mario Castagnino on the occasion of his
75th birthday.}


\vspace{.5cm}
 \centerline{{\bf Abstract}}
\bigskip

\noindent The Aharonov-Bohm effect is a fundamental issue in physics that has been extensively studied in the literature and is discussed in
most of the textbooks in quantum mechanics. The issues at stake are what are the  fundamental  electromagnetic quantities in quantum
physics, if magnetic fields can {\it act at a distance} on charged particles and if the magnetic potentials have a real physical significance.
The Aharonov-Bohm effect  is a very controversial issue. From the experimental side the issues were settled by the remarkable
experiments of Tonomura et al. [Observation of Aharonov-Bohm effect by electron holography, Phys. Rev. Lett. {\bf
48} (1982) 1443-1446 ,  Evidence for Aharonov-Bohm effect with magnetic field completely shielded from electron wave, Phys. Rev. Lett. {\bf 56}
(1986) 792-795]
with toroidal magnets that gave a strong experimental evidence of the physical existence of the Aharonov-Bohm effect, and by
the recent experiment of Caprez et al. [``Macroscopic test of the Aharonov-Bohm effect,"
Phys. Rev. Lett. {\bf 99}  (2007) 210401] that shows that the results of the Tonomura et  al. experiments can not be explained
by the action of a force. Aharonov  and Bohm  [Significance of electromagnetic potentials in the quantum theory,
Phys. Rev. {\bf 115} (1959) 485-491 ] proposed an Ansatz for the solution to the Schr\"odinger equation in simply connected regions of space where there are no
electromagnetic fields. It consists of multiplying  the free evolution by the Dirac magnetic factor. The Aharonov-Bohm Ansatz predicts the results of
the experiments of Tonomura et al. and of Caprez et al.. Recently in [M. Ballesteros, R. Weder, The Aharonov-Bohm effect and Tonomura et al. experiments:
Rigorous results, J. Math. Phys. {\bf 50} (2009) 122108]  we gave the first rigorous proof that the Aharonov-Bohm  Ansatz is
a good approximation to the exact solution for toroidal magnets under the conditions of the experiments of Tonomura et al.. We
provided a rigorous, simple, quantitative, error bound for the difference in  norm between the exact solution and the Aharonov-Bohm Ansatz.
In this paper we prove that these results do not depend  on the particular geometry of the magnets and on the velocities of the incoming
electrons used on the experiments, and on the gaussian shape of the wave packets used to obtain our quantitative error bound.
We consider a general class of magnets that are a finite union of handle bodies.
Each handle body is diffeomorphic to a torus or a ball, and some of them can be patched though the boundary. We formulate the Aharonov-bohm Ansatz
that is appropriate to this general case and  we prove that the exact solution to the Schr\"odinger equation  is given by  the
Aharonov-Bohm Ansatz up to an error bound in norm that is uniform in time and that decays as a constant divided by $v^\rho, 0 < \rho <1$, with
$v$ the velocity. The results of Tonomura et al., of Caprez et al.,  our previous results and the results of this paper give a firm experimental and
theoretical basis to the existence of the Aharonov-Bohm effect and to its quantum nature. Namely, that magnetic fields {\it act at a distance} on charged
 particles, and that this {\it action at a distance} is carried by the circulation of
the magnetic potential what gives a real physical significance to  magnetic potential.


\section{Introduction}
\sss
In classical physics the  dynamics of a charged particle in the presence of a magnetic field is completely described by Newton's equation with the
Lorentz force, $F= q \v \times B$, where  $B$ is the magnetic field, $q$  is the charge of the particle and  $\v$ its velocity. Newton's equation implies
that in classical physics the magnetic field acts locally. If a particle propagates in a region were the magnetic field is zero
the Lorentz force is zero and the trajectory of the particle is a straight line. The dynamics of a classical particle is not affected by
magnetic fields that are located  in regions of space that are not accessible to the particle. The {\it action at a distance} of magnetic fields
on charged particles is not possible in classical electrodynamics. Furthermore, the relevant physical quantity is the magnetic field. The
magnetic potentials have no physical meaning,  they are just a convenient mathematical tool.

In quantum physics this changes in a dramatic way. Quantum mechanics is a Hamiltonian theory were the dynamics of a charged  particle in the presence
of a magnetic field is governed by the equation of Schr\"odinger that  can not be formulated directly in terms of the magnetic field, it requires
the introduction of a magnetic potential. This makes the {\it action at a distance} of magnetic fields possible, since in a region of  space with
non-trivial topology, like the exterior of a torus,  the magnetic potential has to be different from zero if there is a
magnetic flux inside the torus, even if the magnetic field is identically zero outside. The reason is quite simple: if the
magnetic potential is zero outside the torus it follows from Stoke's theorem that the magnetic flux inside  has to be zero. Aharonov and Bohm
observed \cite{ab} that this implies that in quantum physics  the magnetic  flux inside the torus can {\it act
at a distance}  in a charged particle outside the torus, on spite of the fact that the
magnetic field is identically zero along the trajectory of the particle and, furthermore,  that the action of the magnetic field is carried over
by the magnetic potential, what gives  a  real physical significance to  the magnetic potentials.

The possibility that magnetic fields can {\it act at a distance} on
charged  particles and that the magnetic potentials can have a
physical significance is such a strong departure from the physical
intuition coming from classical physics that it is no wonder that
the Aharonov-Bohm effect was, and still is, a very  controversial
issue. In fact, the experimental verification of the Aharonov-Bohm
effect constitutes a test of the validity of the theory of quantum
mechanics itself. For a review of the literature up to 1989 see
\cite{op} and \cite{pt}. In particular, in \cite{pt} there is a
detailed discussion of the large controversy  -involving over three
hundred papers- concerning the existence of the Aharonov-Bohm
effect. For a recent update of this controversy see \cite{to,tn}.

In their seminal paper Aharonov and Bohm \cite{ab} proposed an experiment to verify their theoretical prediction.
They suggested to use a thin straight solenoid. They supposed that the magnetic field
was confined to the solenoid. They suggested to send a coherent electron wave packet towards the solenoid and to split it in two parts,
each one going trough one side of the solenoid, and to bring  both  parts  together behind the solenoid in order  to create an
interference pattern due to the difference in phase in the wave function of each part, produced by the magnetic field  inside the solenoid.
In fact, the existence of this interference pattern was first predicted by Franz \cite{f}.

There is a very large literature for the case of a solenoid both  theoretical
and experimental. The theoretical analysis is reduced to a two
dimensional problem after making the assumption that the solenoid is
infinite. Of course, it is  experimentally  impossible to have an infinite
solenoid. It has to be finite, and the magnetic field has to leak outside. The leakage of the magnetic field was a highly controversial point.
Actually, if we assume that the magnetic field outside the finite
solenoid can be neglected  there is no Aharonov-Bohm  effect at all because, if this is true, the exterior of the finite solenoid is a  simply connected
region of space without magnetic field where the magnetic potential can be  gauged away to zero. In order to circumvent this issue it  was
suggested to use a  toroidal magnet, since it can contain a magnetic field inside without a leak.
The experiments with toroidal magnets where carried over by Tonomura et al. \cite{to3,to1,to2}.
In these remarkable experiments they split a coherent electron wave packet into two parts. One
traveled inside the hole of the magnet and the other outside the magnet. They  bought both parts together behind the magnet and they measured the phase
shift produced by the magnetic flux enclosed in the magnet, giving a
strong evidence of the existence of the Aharonov-Bohm effect. The Tonomura et al. experiments \cite{to3,to1,to2}  are widely
considered  as the only convincing experimental evidence of the existence of
the Aharonov-Bohm effect.

After the fundamental experiments of Tonomura et al.  \cite{to3,to1,to2} the  existence of the Aharonov-Bohm effect was largely accepted and the
controversy shifted into the interpretation of the results of the Tonomura et al. experiments. It was claimed that the outcome of the experiments
could be explained by the action of some force acting on the electron that travels through the hole of the magnet. See, for example, \cite{bo,he} and
the references quoted there. Such a force would accelerate the electron and it would
produce a time delay. In a recent crucial experiment Caprez et al.
\cite{cap} found that the time delay is zero, thus experimentally
excluding the explanation of the results of the Tonomura et al. experiments by the
action of a force.

Aharonov and Bohm \cite{ab} proposed an Ansatz for the solution to the Schr\"odinger equation in simply connected regions of space where there are no
electromagnetic fields. The Aharonov-Bohm Ansatz consists of multiplying  the free evolution by the Dirac magnetic factor \cite{di} (see Definition
\ref{AB-Ansatz} in Section 4). The  Aharonov-Bohm Ansatz predicts the interference fringes observed by  Tonomura et al.  \cite{to3,to1,to2} and it also
predicts the absence of acceleration  observed in the Caprez et al. \cite{cap}  experiments because in the Aharonov-Bohm Ansatz the electron  is not
accelerated since it propagates following the free evolution, with the wave function multiplied by a phase.
As  the experimental issues have already been settled by Tonomura et al. \cite{to3,to1,to2} and by Caprez et al.  \cite{cap}, the whole controversy can
now be summarized in a single mathematical question: is the Aharonov-Bohm  Ansatz a good approximation to the exact solution to the Schr\"odinger equation
for toroidal magnets and under the conditions of the experiments of Tonomura et al. Of course, there have been numerous attempts to give an answer to
this question.
Several Ans\"atze have been provided for the solution to the Schr\"odinger equation and for the scattering matrix, without giving error bound
estimates for the difference, respectively, between the exact solution and the exact scattering matrix, and the Ans\"atze. Most
of these works are qualitative, although some of them give numerical values for their Ans\"atze. Methods like,  Fraunh\"ofer diffraction, first-order
 Born and high-energy approximations, Feynman path integrals and the Kirchhoff method in optics were used to propose the
Ans\"atze. For a review of the literature up to 1989 see \cite{op} and \cite{pt} and for a recent update see \cite{bw},  \cite{bw2}. The lack of any definite
rigorous  result on the validity of the Aharonov-Bohm Ansatz is perhaps  the reason why this controversy lasted for so many years.

It is only very recently that this situation has changed. In our paper \cite{bw2}  we gave the first rigorous proof that the Ansatz of Aharonov-Bohm is
a good approximation to the exact solution of the Schr\"odinger equation. We provided, for the first time, a rigorous quantitative mathematical analysis
 of the Aharonov-Bohm effect with toroidal magnets under the
conditions of the experiments of Tonomura et al. \cite{to3,to1,to2}. We assumed that the incoming free electron  is represented by a gaussian
wave packet, what from
the physical point of view is a reasonable assumption. We provided a rigorous, simple, quantitative, error bound for the difference in
 norm between the exact solution and the approximate solution given by the Aharonov-Bohm Ansatz. Our error bound is uniform in time. We also  proved
 that on the gaussian asymptotic state the scattering operator is given by a constant phase shift, up to a quantitative error bound, that we provided.
 Actually, the error bound is the same in the cases of the exact solution and the scattering operator.

As mentioned above, the  results of \cite{bw2} were proven under the experimental conditions of Tonomura et al., in particular for the  magnets and
 for the velocities  of the incoming electrons considered in  \cite{to3,to1,to2}. This was necessary to obtain rigorous quantitative results
 that can be compared with the  experiments. This raises the question if the experimental results of  \cite{to3,to1,to2} and the rigorous
 mathematical results of \cite{bw2} depend or not on the particular geometry of the magnets, on  the velocities of the incoming electrons
 used in the experiments, and on the gaussian shape of the wave packets.

 In this paper we give a general answer to this question. We assume that the magnet $K$ is a compact submanifold of $\ere^3$.
Moreover, $K= \cup_{j=1}^L K_j$ where $K_j,  1\leq j \leq L$ are the connected components of $K$. We suppose that the $K_j$ are
handlebodies. For a precise definition of handle bodies see \cite{bw}. In intuitive terms, $K$ is the union of a finite number of bodies
diffeomorphic to tori or to balls. Some of them can be patched through the boundary.
See Figure 1.

For the Aharonov-Bohm Ansatz to be valid it is necessary that, to a
good approximation, the electron does not interact with the magnet
$K$, because if the electron hits  $K$ it will be reflected and the
solution can not be the free evolution modified with a phase. This
is true no matter how big the velocity is. Actually, in the case of the infinite solenoid with non-zero cross section this can be seen in
the explicit solution
\cite{ru}. We dealt with this issue
in \cite{bw2} requiring that the variance of the gaussian state be
small in order that the interaction with the magnet was small. In
this paper we consider a general  class of incoming asymptotic
states with the property that under the free classical evolution
they do not hit $K$. The intuition is that for high
velocity the exact quantum mechanical evolution is close to the  free
quantum mechanical evolution and that as the free quantum mechanical
evolution is concentrated on the classical trajectories, we can
expect that, in the leading order for high velocity, we do not see the
influence of  $K$ and that only the influence of the
magnetic flux inside $K$ shows up in the form of a phase, as
predicted by the Aharonov-Bohm Ansatz.

In our general case $K$ has
several holes and the parts of the wave packet that travel through
different holes  adquire different phases. For this reason we
decompose our  electron wave packet into the parts that travel
through the different holes of $K$ and we formulate the
Aharonov-Bohm Ansatz for each one of then. We prove that the exact
solution to the Schr\"odinger equation  is given by  the
Aharonov-Bohm Ansatz up to an error bound in norm that is uniform in time
and that decays as a constant divided by $v^\rho, 0 < \rho <1$, with
$v$ the velocity. In our bound the direction of the velocity is kept
fixed as it absolute value goes to infinite. The results of this
paper complement the results of our previous paper \cite{bw} where
we proved that for the same class of incoming high-velocity
asymptotic states  the  scattering operator is given by
multiplication by a constant phase shift, as predicted by the
Aharonov-Bohm Ansatz.

Our results here, that are obtained with the help of results from  \cite{bw}, prove in a
qualitative way that the Ansatz of Aharonov-Bohm is a good
approximation to the exact solution of the Schr\"odinger equation
for high velocity for a very general class of magnets $K$ and of
incoming asymptotic states, proving that the experimental results of
Tonomura et al. \cite{to3,to1,to2} and of Caprez et al. \cite{cap}
and the rigorous mathematical results of  \cite{bw2}  hold in
general and that they do not   depend on the particular geometry
of the magnets, on  the velocities of the incoming electrons used in the experiments, and on
the gaussian shape of the wave packets.

Summing up, the experiments of Tonomura et al. \cite{to3,to1,to2}
give a strong  evidence of the existence of the interference fringes
predicted by Franz \cite{f} and by Aharonov and Bohm \cite{ab}. The
experiment of Caprez et al. \cite{cap} verifies that the
interference fringes are not due to a force acting on the electron,
and the results \cite{bw}, \cite{bw2} and on this paper
rigorously prove that quantum mechanics theoretically predicts the
observations of these experiments  in a extremely precise
quantitative way under the experimental conditions in \cite{bw2} and
in a qualitative way for general magnets  and incoming asymptotic
states   on \cite{bw} and on this paper.  These results  give a firm
experimental and theoretical basis to the existence of the
Aharonov-Bohm effect \cite{ab} and to its quantum nature. Namely,
that magnetic fields {\it act at a distance} on charged particles, even if
they are identically zero in the space accessible to the particles,
and that this action at a distance is carried by the circulation of
the magnetic potential, what gives magnetic potentials a real
physical significance.

The results of this paper, as well as the ones of \cite{bw}, \cite{bw2}, and of \cite{n}, \cite{w1}  where the Aharonov-Bohm effect in the case of
solenoids contained inside infinite cylinders  with arbitrary cross section was rigorously studied,  are proven using  the method introduced in \cite{ew} to estimate the
high-velocity limit of solutions to the Schr\"odinger equation  and of the scattering operator.

The paper is organized as follows. In Section 2 we state preliminary results that we need. In Section 3 we obtain estimates in norm for the leading
order at high velocity of the exact solution to the Schr\"odinger equation in the case where besides the magnetic flux inside $K$ there are a magnetic
field and an electric potential outside $K$. Our estimates are uniform in time. These results are of independent interest and they go beyond the
Aharonov-Bohm effect. The main results of this section are Theorems \ref{theor-3.9} and \ref{theor-3.10} and  Section 3.2
where the physical interpretation of our estimates is given. In Section 4 we consider the Aharonov-Bohm effect and we prove our estimates that
show that the Aharonov-Bohm Ansatz is a good approximation to the exact solution to the Schr\"odinger equation. The main results are Theorems
\ref{theor-4.12}, \ref{theor-4.13} and \ref{theor-4.14}. In the Appendix we prove a result that we need, namely the triviality of the first group of
singular homology of the sets where electrons that travel through  different holes are located.

Let us mention some related rigorous results on the Aharonov-Bohm effect. For further references see \cite{bw} \cite{bw2}, and \cite{w1}.
In \cite{hel}, a semi-classical analysis of the Aharonov-Bohm effect in bound-states in two dimensions is given. The papers \cite{rou},
\cite{ry1}, \cite{ya1}, and \cite{ya2} study the scattering matrix
for potentials of Aharonov-Bohm type in the whole space.

Finally some words about our notations and definitions. We
denote by $C$ any finite positive constant whose value is not
specified. For any $ x\in \ere^3, x \neq 0$, we denote, $\hat{x}:=
x/|x|$.  for any $ \v \in \ere^3$ we designate, $ v:= |\v|$.  By
$B_R(x)$ we denote the open ball of center $x$ and radius $R$. $B_R(0)$ is denoted by $B_R$.
For any set $O$ we denote by $F(x \in O)$ the operator of
multiplication by the characteristic function of $O$. By $\|\cdot
\|$ we denote the norm in $L^2(\Lambda)$ where, $\Lambda:= \ere^3 \setminus K$. The norm of $L^2(\ere^3)$
is denoted by $\|\cdot \|_{\ds L^2(\ere^3)}$.  For any open set,
$O$, we denote by $\mathcal H_s(O), \, s=1,2,\cdots$  the Sobolev
spaces \cite{ad} and by $\mathcal H_{s,0}(O)$ the closure of
$C^\infty_0(O)$ in the norm of $\mathcal H_{s}(O)$. By  $\mathcal{B}(O)$ we designate the Banach space of all bounded operators on $L^2(O)$.

We use notions of homology and cohomology as defined, for example, in \cite{br},  \cite{dr}, \cite{gh}, \cite{h},
 and \cite{w}. In particular, for a set $ O \subset \ere^3$ we denote by $H_1(O; \ere)$ the first group of singular homology with coefficients
 in $\ere$, \cite{br} page 47, and by $H^1_{de R}(O)$ the first de Rham cohomology class of $O$ \cite{w}.

We define the Fourier transform as a unitary operator on $L^2(\ere^3)$ as follows,
$$
\hat {\phi}(p):= F \phi(p):= \frac{1}{(2 \pi)^{3/2}} \int_{\ere^3} e^{-i p\cdot x} \phi (x)\, dx.
$$

We define functions of the operator $\mo:=-i \nabla$ by Fourier transform,
$$
f(\mo) \phi:= F^\ast f(p) F \phi, \, D(f(\mo)):= \{ \phi \in L^2(\ere^3): f(p) \,\hat{\phi}(p) \in L^2(\ere^3)
\},
$$
for every measurable function $f$.

\section{Preliminary Results}
\sss We study the propagation of a non-relativistic  particle -an
electron for example- outside a bounded  magnet, $K$, in three
dimensions, i.e. the electron  propagates in  the  exterior domain
$\Lambda := \ere^3 \setminus K$. We asssume that inside $K$ there is
a magnetic field that produces a magnetic flux. We suppose, furthermore, that in $\Lambda$ there are an
electric potential $V$ and a magnetic field $B$. This is a more general situation than the one of the Aharonov-Bohm effect.

\subsection{The Magnet $K$}
We assume that the magnet $K$ is a compact submanifold of $\ere^3$.
Moreover, $K= \cup_{j=1}^L K_j$ where $K_j,  1\leq j \leq L$ are the
connected components of $K$. We suppose that the $K_j$ are
handle bodies. For a precise definition of handlebodies see
\cite{bw} were we study in detail the homology and the cohomology of
$K$ and  $\Lambda$. In intuitive terms, $K$ is the union of a finite
number of bodies diffeomorphic to tori or to balls. Some of them can
be patched through the boundary. See Figure 1.
\subsection{The Magnetic Field and the Electric Potential}
In the following assumptions we summarize the conditions on the magnetic field and the electric potential that
we use (see \cite{bw}). We denote by $\Delta$ the self-adjoint realization of the Laplacian in $L^2(\ere^3)$ with domain
$\mathcal  H_2(\ere^3)$. Below we assume that $V$ is $\Delta$- bounded with relative bound zero. By this we
mean that the extension of $V$ to $\ere^3$ by zero is $\Delta-$ bounded with relative bound zero.
Using a extension operator from $\mathcal H_2(\Lambda)$ to $H_2(\ere^3)$ \cite{tr} we prove that this is
equivalent to require that $V$ is  bounded from $\mathcal H_2(\Lambda)$ into $L^2(\Lambda)$ with relative bound
zero.
We denote by $\| \cdot \|_{\ds \mathcal{B}(\ere^3)}$ the operator norm in $L^2(\ere^3)$.

\begin{assumption}\label{ass-2.1}
{ \rm We assume that the magnetic field,
 $B$, is a real-valued, bounded $2-$ form in $ \overline{\Lambda}$,
that is continuous in a neighborhood of $\partial K$, and furthermore,
\begin{enumerate}
                                                                                                                                                                                                                                                                                                                                                                                                                                                                                                                                                                                                                                                                                                                                                                                                                                                                                                                                                                                                                                                                                                                                                                                                                                                                                                                                                                                                                                                                                                                                                                                                                                                                                                                                                                                                                                                                                                                                                                                                                                                                                                                                                                                                                                                                        \item
$ B \, \hbox{\rm is closed}:    d B|_{\Lambda} \equiv \,\hbox{\rm div} B=0$.

\item
There are no magnetic monopoles in $K$:
\beq
\int_{\partial K_{j}} B=0, \, j \in \{1,2,\cdots,L\}.
\label{2.1}
\ene

\item
\beq
\left|B(x)\right| \leq C (1+|x|)^{- \mu}, \, \hbox{\rm for some}\,\, \mu > 2.
\label{2.2}
\ene
\item

  $d* B|_{\Lambda} \equiv \hbox{\rm curl}\, B $ is  bounded and,
\beq
|\hbox{\rm curl}\, B |  \leq C (1+|x|)^{- \mu}.
\label{2.3}
\ene
\item
The electric  potential, $V$, is a real-valued function, it is  $\Delta-$bounded, with relative bound zero and
\beq
\left\|   F(|x|\geq r) V(-\Delta+I)^{-1}\right\|_{\ds \mathcal{B}(\ere^3)}  \leq C (1+ r)^{- \alpha}, \,\hbox{\rm for some}\, \alpha >1 .
\label{2.4}
\ene
\end{enumerate}}
\end{assumption}
Condition (\ref{2.1}) means that the total contribution of magnetic
monopoles inside each component $ K_j $ of  the magnet is $ 0 $. In
a formal way we can use Stokes theorem to conclude that
$$
\int_{\partial K_j} B =0 \Longleftrightarrow \int_{K_j} \, \hbox{\rm div}\, B=0,j \in \{1,2,\cdots,L\}.
$$
As $\hbox{\rm div}\, B$ is the density of magnetic charge, $\int_{\partial K_j} B$ is the total magnetic charge
inside $K_j$, and our condition (\ref{2.1}) means that the total magnetic charge inside $K_j$ is zero.  This condition in fulfilled if there is no
magnetic monopole inside $K_j, j \in \{1,2,\cdots,L\}$.

 Furthermore, condition
(\ref{2.4}) is equivalent to the following assumption \cite{rs3}

\beq
\left\|V (-\Delta+I)^{-1}  F(|x|\geq r)  \right\|_{\ds \mathcal{B}(\ere^3)}  \leq C (1+r)^{- \alpha}, \,\hbox{\rm for some}\, \alpha >1 .
\label{2.5}
\ene
Condition (\ref{2.4}) has a clear intuitive meaning, it is
 a condition on   the decay of $V$ at infinity.
However, in the proofs below we use the equivalent statement (\ref{2.5}).

\subsection{The Magnetic Potentials}

Let  $\{ \hat{\gamma}_j\}_{j=1}^m$  be the closed curves defined in  equation   (2.6) of \cite{bw} (see Figure 1). We prove in  Corollary 2.4 of
\cite{bw} that the equivalence classes of these curves
are a basis of the first singular homology group of $\Lambda$. We introduce below a function
that gives the magnetic flux across  surfaces that have  $\{\hat{\gamma}_j\}_{j=1}^m$ as their boundaries.

\begin{definition} \label{def-2.2}{\rm
The flux, $\Phi$, is a function $\Phi: \{\hat{\gamma}_j\}_{j=1}^m \rightarrow \ere$.}
\end{definition}
We now define a class of magnetic potentials with a given flux modulo $2 \pi$.

\begin{definition} \label{def-2.3}{\rm
Let $B$ be a closed $2-$ form that satisfies  Assumption \ref{ass-2.1}. We denote by
 $\p2$ the set of all continuous $1-$ forms, $A$, in $ \overline{\Lambda}$ that satisfy.
\begin{enumerate}
\item
\beq
|A(x)| \leq C \frac{1}{1+|x|},
\label{2.6}
\ene
\beq
\left|A (x)\cdot \hat{x}\right| \leq C (1+|x|)^{- \beta_l}, \quad
\beta_l > 1, \,\hbox{\rm where}\, \hat{x}:= x/ |x|.
\label{2.7}
\ene
\item
\beq
\int_{\hat{\gamma_j}}\, \ap= \Phi (\hat{\gamma_j})+ 2 \pi n_j(A), n_j(A) \in \ZETA,    \, j \in \{1,2,\cdots, m \}.
\label{2.8}
\ene
\item
\beq
d A|_{\Lambda}\equiv \, \curl\, A= B|_{\Lambda}.
\label{2.9}
\ene
\end{enumerate}
Furthermore,  we say that two potentials, $A, \ta \in \p2$ have the same fluxes if
\beq
\int_{\hat{\gamma_j}}\, A = \int_{\hat{\gamma_j}}\, \ta,    \, j \in \{1,2,\cdots, m \}.
\label{2.10}
\ene
Moreover, we say that $ A \in \p2$ is short range if
\beq
|A(x)| \leq C \frac{1}{(1+|x|)^{\beta}}, \quad \beta >1.
\label{2.10b}
\ene
We denote by $\sp2$ the set of all potentials
in $\p2$ that are short range.}
\end{definition}

The definition of  the flux $\Phi$ depends on the particular choice of the curves $\{\hat{\gamma}_j\}_{j=1}^m$.
 However, the class $\p2$ is independent of this particular choice. In fact it can be equivalently  defined taking any other basis
 of the first singular homology group in $\Lambda$. See \cite{bw}.
By Stoke's theorem the circulation $\int_{\hat{\gamma}_{j}} A$ of a potential  $ A \in \p2$ represents the flux
 of the magnetic field $B$ in any surface whose boundary is $ \hat{\gamma}_j, j= 1,2,\cdots,m$. As the magnetic
field is {\it a priori} known outside the magnet, it is natural to
specify the magnetic potentials fixing fluxes of the magnetic field
in surfaces inside the magnet taking the circulation of $A$ in
closed curves in the boundary of $K$. We prove in \cite{bw} that
this gives the same class of potentials. We find, however, that it
is technically more convenient to work with closed curves in
$\Lambda$ that define a basis of the first singular homology group.
 Note that in \cite{bw} we use the same symbol to denote a larger class of
 magnetic potentials where (\ref{2.7}) is only required to hold $L^1$ sense. Here we assume that it holds in pointwise sense to obtain precise
 error bounds.

In theorem 3.7 of \cite{bw} we construct the Coulomb potential, $A_C$, that belongs to $\sp2$ with   $n_j(A)=0, \, j \in \{1,2,\cdots, m\}$.
For this purpose condition (\ref{2.1}) is essential.

In  Lemma 3.8 of \cite{bw} we prove that for any $A, \ta \in \p2$ with the same fluxes there is a gauge transformation between them. Namely, that there
is there is a $C^1 \, 0-$ form $\lambda$ in $\overline{\Lambda}$
such that,
\beq
\tilde{A} - A= d \lambda.
\label{2.11}
\ene
Moreover, we can take $\lambda(x):=\int_{C(x_0,x)}(\tilde{A}-A)$ where $x_0 $
is any fixed
point in $\Lambda$ and $C(x_0,x)$ is any
 curve from $x_0$ to $x$. Furthermore, $\lambda_\infty(x):=\lim_{r\rightarrow \infty} \lambda(rx)$ exists
 and it is continuous in $  \mathbb{R}^3 \setminus \{ 0 \}  $ and homogeneous of order zero, i.e.
  $\lambda_\infty(r x)=\lambda_\infty(x), r >0,
  x \in  \ere^3 \setminus\{0\}$. Moreover,
\beq
\begin{array}{c}
|\lambda_\infty(x)-\lambda(x)|\leq \int_{|x|}^\infty \, b(|x|), \hbox{\rm for some}\,\, b(r)\in L^1(0,\infty),
\, \\\\ \hbox{\rm and} \,\,  |\lambda_\infty (x+y)-\lambda_\infty(x)|\leq C |y|, \forall x: |x|=1, \,
 {\rm and}\, \forall y: |y| < 1/2.
\end{array}
\label{2.12}
\ene

\subsection{The Hamiltonian}

\noindent Let us denote  $\mo:=-i\nabla$.
The Schr\"odinger equation for an electron in $\Lambda$  with electric potential $\mathbf V$ and magnetic
field $\mathbf B$   is given by
\beq
i\hbar \frac{\partial}{\partial t} \phi = \frac{1}{2 M} ({\mathbf P}- \frac{q}{ c}\mathbf A)^2+q \,\mathbf V,
\label{2.13}
\ene
where $\hbar$ is Planck's constant, ${\mathbf P}:=\hbar \mo$ is the momentum operator, $c$ is the
speed of light, $M$ and $q$ are,
respectively, the mass and the charge of the electron and $ \mathbf A$ a magnetic potential with
$\hbox{\rm curl} \mathbf A= \mathbf B$. To simplify the notation
we multiply both sides of (\ref{2.12}) by $\frac{1}{\hbar}$ and we write Schr\"odinger's equation as follows

\beq
i \frac{\partial} {\partial t} \phi= \frac{1}{2m} (\mo- A)^2 \phi+V\phi,
\label{2.14}
\ene
with $m:= M/ \hbar, A= \frac{q}{\hbar c}\mathbf A$ and $V:= \frac{q}{\hbar} \mathbf V$.
Note that since we write Schr\"odinger's equation in this form our Hamiltonian below is the physical
Hamiltonian divided by $\hbar$. We fix the flux modulo $2\pi$ by taking $A \in \mathcal A _{\Phi, 2 \pi}$ where $B:=\frac{q}{\hbar c}
{\mathbf B}$. Note that this corresponds
to fixing the circulations of $\mathbf A$ modulo $\frac{\hbar c}{q} 2 \pi$, or equivalently, to fixing
the fluxes
of the magnetic field $\mathbf B$ modulo  $\frac{\hbar c}{q} 2 \pi$.

We define the quadratic form,
\beq
h_0(\phi,\psi):= \frac{1}{2 m}(\mo \phi,\mo \psi), \, D(h_0):= \mathcal H_{1,0}(\Lambda).
\label{2.15}
\ene
The associated positive operator in $L^2(\Lambda)$  \cite{ka}, \cite{rs2} is $\frac{-1}{2m} \Delta_D$
where $\Delta_D$ is  the  Laplacian with
Dirichlet boundary condition on $\partial \Lambda$. Note that the functions in  $\mathcal H_{s,0}(O)$ vanish in trace sense in the boundary of $O$.
We define
$H(0,0):= \frac{-1}{2m} \Delta_D$.
By elliptic regularity \cite{ag}, $D(H(0,0))= \mathcal H_2(\Lambda) \cap \mathcal H_{1,0}(\Lambda)$.

For any $ A \in \p2$ we define,
\beq
\begin{array}{c}
h_A(\phi,\psi):= \frac{1}{2m}\left( (\mo-A)\phi, (\mo-A) \psi\right)=h_0(\phi,\psi)
+ \frac{1}{2m}(-(\mo \phi,A\psi)-(A\phi,\mo\psi)) + \frac{1}{2m} (A \phi, A \psi ),\\
 D(h_A)= \mathcal H_{1,0}(\Lambda).
\end{array}
\label{2.16}
\ene
As the quadratic form $ -\frac{1}{2m} ((\mo \phi,A\psi)+(A\phi,\mo\psi))+ \frac{1}{2m} (A \phi, A \psi )$ is $h_0-$ bounded with relative bound zero,
$h_A$ is closed and positive. We denote by $H(A,0)$ the associated positive self-adjoint operator \cite{ka},
\cite{rs2}. $H(A,0)$ is the Hamiltonian with magnetic potential $A$.
As  the electric potential $V$  is $h_0-$ bounded with relative bound zero it follows \cite{ka}, \cite{rs2} that the quadratic form,
\beq
h_{A,V}(\phi, \psi):= h_A( \phi,\psi)+(V \phi,\psi), \,
 D(h_{A,V})= \mathcal H_{1,0}(\Lambda),
\label{2.17}
\ene
is closed and bounded from below. The associated operator, $H(A,V)$, is self-adjoint and bounded from below.
$H(A,V)$ is the Hamiltonian with magnetic potential $A$ and electric potential $V$.

Suppose that $ \div A$ is bounded. In this case the operator $
\frac{1}{2m} (-2 A\cdot \mo- (\mo\cdot A)   +A^2 )$ is $H(0, 0)$
bounded with relative bound zero and we have that $H(0,0)-\frac{1}{2m} ( 2 A\cdot \mo+ (\mo\cdot A))+
\frac{1}{2m}A^2$ is self-adjoint on the domain of $H(0,0)$ and since
also $V$ is $H(0,0)$ bounded with relative bound zero we have that,
 \beq
  H(A,V)= H(0,0)-\frac{1}{2m} (2 A\cdot \mo+ (\mo \cdot
A))+\frac{1}{2m}A^2+V, \, D(H(A,V))= \mathcal H_2(\Lambda)\cap
\mathcal H_{1,0}(\Lambda).
\label{2.18}
\ene
We define the Hamiltonian $H(A,V)$  in $L^2(\Lambda)$  with Dirichlet boundary
condition at $\partial \Lambda$, i.e. $\psi=0$ for $x \in \partial
\Lambda$. This is the standard boundary condition that corresponds
to an impenetrable magnet $K$. It implies that the probability that
the electron is at the boundary of the magnet is zero.  Note that
the Dirichlet boundary condition is invariant under gauge
transformations. In the case of the impenetrable magnet  the
existence of the Aharonov-Bohm effect is more striking, because in
this situation there is zero interaction of the electron with the
magnetic field inside the magnet. Note, however, that once a
magnetic potential is chosen the particular self-adjoint boundary
condition  taken at $\partial \Lambda$ does not play an essential
role in our calculations.  Furthermore, our results hold also for a
penetrable magnet where the interacting Schr\"odinger equation  is
defined in all space. Actually, this later case is slightly simpler
because we do not need to work with two Hilbert spaces,
$L^2(\ere^3)$ for the free evolution, and $L^2(\Lambda)$ for the
interacting evolution, what simplifies the proofs. We prove in
Theorem 4.1 of \cite{bw} that if $A,\ta \in \p2$ the Hamiltonians $
H(A,V)$ and $H(\ta,V)$ are unitarily equivalent and we give
explicitly the unitary operator that relates them.

\subsection{The Wave and Scattering Operators}

Let $J$ be the identification operator from $L^2(\ere^3)$ onto $L^2(\Lambda)$ given by multiplication by the
characteristic function of $\Lambda$. The wave operators are defined as follows,

\beq
 W_{\pm}(A,V):= \hbox{\rm s-}\lim_{t \rightarrow \pm \infty} e^{it
H(A,V)}\, J\, e^{-it H_0},
\label{2.19}
\ene
provided that the strong limits exist.  We prove in \cite{bw}  that if  Assumption
\ref{ass-2.1} holds the wave operators exist and are partially
isometric for every $A\in \p2$ and that $J$ can be replaced by the
operator of multiplication by any function $\chi \in
C^\infty(\ere^3)$ that satisfies  $\chi(x)=0$ in a neighborhood of
$K$ and $\chi (x)=1$ for $ x\in \ere^3\setminus  B_{R}$  where $K \subset B_R$. Furthermore, the wave operators satisfy the intertwining relations,

\beq
e^{it H(A,V)} \, W_{\pm}(A,V) = W_{\pm}(A,V)\,e^{it H_0}.
\label{2.20}
\ene

 Moreover, if $ A,\ta \in \p2$ and they have the same fluxes (for the case where the fluxes are not equal see \cite{bw})
\beq
W_\pm(\tilde{A},V)=  e^{i \lambda(x)}\, W_\pm(A,V)\, e^{-i \lambda_\infty(\pm\mo)}.
\label{2.21}
\ene
The scattering operator is defined as
\beq
S(A,V):= W_+^\ast(A,V)\, W_-(A,V).
\label{2.22}
\ene
If $A,\ta \in \p2$ \cite{bw},
\beq
S(\tilde{A},V)= e^{i \lambda_\infty(\mo)}\, S(A, V)\, e^{-i \lambda_\infty(-\mo)}, \, \tilde{A}, A  \in \p2.
\label{2.23}
\ene
If $A, \ta \in \sp2$ (or more generally if $A-\ta$ satisfies (\ref{2.10b}))  $\lambda_\infty$ is constant and  by (\ref{2.23})
$S(\tilde{A},V)= S(A,V)$. That is to say, the scattering operator is  uniquely defined  by $K,B,V$ and the flux $\Phi $ modulo $2\pi$, if we restrict
the potentials to be of short range.

\section{Uniform Estimates}
\sss
We first prepare some results that we need.

In Theorem 3.2 of \cite{bw} we proved that $B$ has an extension to a closed 2-form in $\ere^3$. Below we use the same symbol, $B$, for this closed
extension. Furthermore, in Theorem 3.7 of \cite{bw} we constructed the Coulomb potential, $A_C \in \sp2$, that actually has the fluxes (\ref{2.8})
with $n_j(A)=0, \, j \in \{1,2,\cdots, m\}$. In fact, $A_C$ extends to a continuous 1-form in $\ere^3$, that we denote by the same symbol, $A_C$, such that
$\div A_C$ is infinitely differentiable and with support contained in $K$. See the proof of Lemma 5.6 of \cite{bw}.
For any potential $A \in \p2$ we can construct a Coulomb potential $A_C$ with the same fluxes as $A$. As mentioned above (see (\ref{2.11})),
by Lemma 3.8 of \cite{bw}
there is a $C^1\, 0-$ form  $\lambda$  such that
\beq
A= A_C + d \lambda.
\label{3.1}
\ene
Note that  $\lambda$ has an extension to a $C^1\, 0-$ form in $\ere^3$ ( Theorem 4.22, p.311 \cite{tr} ) that we denote by the same symbol, $\lambda$.
Then, equation (\ref{3.1}) defines an extension of $A$ to a continuous one form in $\ere^3$ that we denote by the same symbol, $A$. Furthermore, the
gauge transformation formula (\ref{2.11}) holds for the extensions of $\ta, A$ and $\lambda$ to $\ere^3$.

We define for $\v \in \ere^3 \setminus 0$,
\beq
\eta (x,t):= \int_0^t (\hv\times B)(x+\tau\hv)d\tau,
\label{3.2}
\ene
\beq
L_{A,\hv}(t):= \int_0^t\, \hv\cdot A(x+\tau \hv) d\tau, -\infty \leq t \leq \infty,
\label{3.3}
\ene
\beq
b(x,t):= A(x+t\hv)+\int_0^t (\hv\times B)(x+\tau\hv)d\tau.
\label{3.4}
\ene
For ${\mathbf f}: \ere^3\times \ere \rightarrow \ere^3$ with $ \mathbf f_t(x):={\mathbf f}(x,t) \in L^1_{\hbox{\rm loc}}
(\ere^3,\ere^3)$ we define,
\beq
\Xi_{\mathbf f}(x,t):= \frac{1}{2m}\chi (x)\left [
-\mo\cdot \mathbf f(x,t)-\mathbf f(x,t)\cdot \mo
+(\mathbf f(x,t))^2\right],
\label{3.5}
\ene
where  $\chi \in C^{\infty}(\ere^3)$  satisfies $\chi(x)= 0$ for  $x$ in a neighborhood of $K$, $\chi(x)=1, x \in \{ x: |x| \geq R \}$ with $R$ such
that $ K \subset B_R$.

It follows by Fourier transform that under translation in configuration or momentum space generated, respectively, by $\mo$ and $x$
we obtain

\beq \label{3.6}
e^{i \mo \cdot \v t} \, f(x) \, e^{-i \mo \cdot \v t}= f(x+\v t),
\ene
\beq \label{3.7}
e^{-im\v\cdot x}\, f(\mo )\, e^{im\v\cdot x}= f(\mo +m \v ),
\ene
and, in particular,
\beq
e^{-im\v\cdot x}\, e^{-it H_0 } \, e^{im\v\cdot x}= e^{-i mv^2 t/2}\, e^{-i\mo\cdot \v t}\,
 e^{-i tH_0}.
\label{3.8}
\ene
We define \cite{w1},
\beq
\label{3.9}
 H_1:= \frac{1}{v} e^{-im\v\cdot x}\, H_0 \,e^{im\v\cdot x},\,\,  H_2:= \frac{1}{v}\, e^{-im\v\cdot x}H(A,V)
 \, e^{im\v\cdot x}.
\ene
We need  the following lemma from \cite{w1}.
\begin{lemma} \label{lemm-3.1}
For any $f \in C^\infty_0(B_{\eta})$ and for any $j=1,2,\cdots$ there is a
constant
$C_j$ such that
\beq
\left\| F\left(|x- z| > \frac{|z|}{4}\right) e^{-i\frac{z}{v} H_0} \, f\left(\ds \frac{\mo-m \v}{\sqrt{v}}\right)\,
F\left( |x| \leq |z|/8\right)\right\|_{{\mathcal B}(\ere^3)} \leq
C_j (1+|z|)^{-j},
\label{3.10}
\ene
for $v:=|\v| > (8\eta/ m)^{2}$.
\end{lemma}

\noindent {\it Proof:} Corollary 2.2 of \cite{w1} with $Q=0$. Note that the proof in three dimensions is the same as the one in two dimensions given in
\cite{w1}.

\begin{lemma} \label{lemm-3.2}
Let $g \in C^\infty_0(\ere^3)$ satisfy, $ g(p)=1, |p| < m/16$ and $ g(p)=0, |p|  \geq  m/8$.
Suppose that $V$ satisfies (\ref{2.4}) or, equivalently, (\ref{2.5}). Then, for any compact set $ D \subset \ere^3$ there is a constant $C$ such that
\beq
\| V e^{ -i z H_1}  g\left(\ds \frac{\mo}{\sqrt{v}}\right)  \phi \|_{L^2(\ere^3)} \leq C (1+|z|)^{-\alpha} \, \|\phi\|_{ \mathcal H_2(\ere^3)},
\label{3.11}
\ene
for all $v >1, z \in \ere$ and all     $\phi \in  \mathcal H_2(\ere^3)$ with support in $ D$. Furthermore, if  $ V \in L^\infty(\ere^3)$ and
 for some $ z\in \ere$,
\beq
\left\|V(x) F(|x-z \hv|\leq |z/4|)\right\|_{ \mathcal{B}(\ere^3)} \leq C (1+|z|)^{-\alpha},\quad \forall x \in \ere^3,
\label{3.12}
\ene
 then, there is a constant $C_1$ such that
\beq
\| V e^{ -i z H_1}  g\left(\ds \frac{\mo}{\sqrt{v}}\right)\phi\|_{L^2(\ere^3)} \leq C_1 (1+|z|)^{-\alpha} \, \|\phi\|_{ L^2(\ere^3)},
\label{3.13}
\ene
for all $v>1$ and all $\phi \in L^2(\ere^3)$ with support in $ D$. The constant $C_1$ depends only on $\| V\|_{L^\infty}$ and on $C$.
\end{lemma}

\noindent {\it Proof:} By (\ref{3.8}),
\beq
\begin{array}{l}
\left\| V e^{ -i z H_1} g\left(\ds \frac{\mo}{\sqrt{v}}\right) \phi\right\|_{L^2(\ere^3)} \leq
 \left\| V (-\Delta+1)^{-1} F(|x-z \hv| > |z|/4)\,e^{ -i \frac{z}{v} H_0} g\left(\ds \frac{\mo-m \v}{\sqrt{v}}\right)\, F(|x|\leq |z|/8 )
 \right\|_{ \mathcal{B}(\ere^3)}
 \\\\
  \|\phi\|_{{\mathcal H}_2(\ere^3)}+
 \left\| V (-\Delta+1)^{-1} F(|x-z \hv| \leq |z|/4)\right\|_{ \mathcal{B}(\ere^3)}  \, \|\phi\|_{{\mathcal H}_2(\ere^3)}+  \| F(|x|> |z|/8) (-\Delta+1)\phi\|_{L^2(\ere^3)}.
\end{array}
\label{3.14}
\ene
Equation (\ref{3.11}) follows from (\ref{2.5}, \ref{3.10}, \ref{3.14}) and using that as  $\phi$ has compact support in $D$,
$$
\| F(|x|> |z|/8) (-\Delta+1)\phi\|_{L^2(\ere^3)} \leq C_l (1+|z|)^{-l} \| (1+|x|)^l (\Delta+1)\phi \|_{L^2(\ere^3)} \leq
C_l\,(1+|z|)^{-l} \, \|\phi\|_{ \mathcal H_2(\ere^3)}.
$$
Equation (\ref{3.12}) is proven in
the same way, but as the regularization $(-\Delta+1)^{-1}$ is not needed we  obtain the norm of $\phi$ in $L^2(\ere^3)$.

\bull

With  $g$ as in Lemma \ref{lemm-3.2} we denote,
\beq
\tilde{\phi}:= g(\mo /\sqrt{v}) \, \phi, \quad v>0.
\label{3.15}
\ene
By Fourier transform we prove that,
\beq
\left\| \tilde{\phi}-\phi \right\|_{L^2(\ere^3)}\leq  \frac{C}{1+v}\,\|\phi\|_{ \mathcal H_2(\ere^3)}.
\label{3.16}
\ene
\subsection{High-Velocity  Solutions to the Schr\"odinger Equation}
At the time of emission, i.e., as  $t \rightarrow -\infty$, electron wave packet is
far away $K$  and  it does not  interact with it, therefore, it can be parametrised with kinematical variables and
it can be assumed that it follows the free evolution,

\beq \label{3.17}
i \frac{\partial}{\partial t} \phi (x,t) =
H_0 \phi(x,t), x \in  \ere^3, t \in \ere. \ene where $H_0$ is the
free Hamiltonian.
\beq
\label{3.18}
H_0 : =  \frac{1}{2 m} {\mo
}^2.
\ene
We represent the emitted electron wave packet  by the free evolution of an asymptotic state with velocity $\v$,

\beq
\label{3.19}
\varphi_{\v}:= e^{i m \v\cdot x}\,\varphi_0, \quad \varphi_0 \in L^2(\ere^3).
\ene
Recall that in the momentum representation $ e^{im \v \cdot x}$ is a translation operator by the vector $ m \v
 $, what implies that
the asymptotic state (\ref{3.19}) is centered at the classical momentum $m \v $ in the momentum
representation,
$$
\hat{\varphi}_{\v}(p)= \hat{\varphi}_0(p-m\v).
$$
Then, the electron wave packet  is represented at the time of emission by the following incoming
wave packet that is a solution to the free Schr\"odinger equation (\ref{3.17})
\beq
\psi_{\v,0}:= e^{-itH_0} \, \varphi_\v.
\label{3.20}
\ene
The (exact) electron wave packet, $\psi_{\v}(x,t)$, satisfies the interacting Schr\"odinger equation  (\ref{2.14}) for all times and as
$t \rightarrow -\infty$ it has to approach the incoming wave packet, i.e.,
$$
\lim_{t \rightarrow -\infty}  \left\| \psi_{\v}-J \psi_{\v,0}\right\|=0.
$$
Hence, we have to solve the interacting Schr\"odinger equation (\ref{2.14}) with initial
conditions at minus infinity. This is accomplished with  wave operator $ W_{-} $. In fact, we have that,
\beq
\psi_{\v}= e^{-it H(A,V)}\, W_-(A,V)\, \varphi_{\v},
\label{3.21}
\ene
because, as $ e^{-it H(A,V) }$ is unitary,
$$
\lim_{t \rightarrow - \infty}\left\|e^{-i t H(A,V) } \,W_{-}\, \varphi_{\v}- J \, e^{-it H_0 } \varphi_{\v} \right\|=0.
$$
Moreover,
\beq
\lim_{t \rightarrow  \infty}\left\|e^{-it H(A,V)} \,W_{-}\,\varphi_{\v} - J \, e^{-it H_0 } \varphi_{\v,+}\right\|=0,\quad
\mathrm{where} \, \varphi_{\v,+}:= W_+^\ast W_- \, \varphi_{\v}.
\label{3.22}
\ene
This means that -as to be expected- for large positive times, when the exact electron wave packet is far away from  $K$, it behaves as the
outgoing solution to the free Schr\"odinger equation (\ref{3.17})
\beq
 e^{-i tH_0 } \varphi_{\v,+},
\label{3.23}
\ene
where the Cauchy data at $t=0$ of the incoming and the outgoing   wave packets (\ref{3.19}, \ref{3.23}) are related by the scattering operator,

$$
\varphi_{\v,+} = S(A,V) \, \varphi_{\v}.
$$
 In order to see the Aharonov-Bohm effect we need to separate the
effect of $K$ as a rigid body  from that of the magnetic flux inside
$K$. For this purpose we need asymptotic states that have negligible
interaction with $K$ for all times. This is possible if the velocity
is high enough, as we explain below.

For any $\v \neq 0$ we denote,
\beq
\Lambda_{\hv}:= \{x \in \Lambda: x+\tau \hv \in \Lambda,\, \forall \tau \in \ere\}.
\label{3.24}
\ene
Let us consider  asymptotic states (\ref{3.19})
where  $ \varphi_0 $ has compact support contained in $\Lambda_{\hv}$. For the discussion below it is  better to parametrise the free evolution
of $\varphi_{\v}$ by the distance $z=v t$ rather than by the time $t$. At distance $z$ the state is given by,
\beq
e^{ -i \frac{z}{v} H_0}\, \varphi_{\v}= e^{im\v\cdot x}\,   e^{-i\frac{mzv}{2}}\, e^{ -i \frac{z}{v} H_0}\, e^{-i\mo\cdot z \hv}  \varphi_{0},
\label{3.25}
\ene
where we used (\ref{3.8}). Note that $e^{-i\mo\cdot z \hv}$ is a translation in straight lines along the classical free evolution,
\beq
\left( e^{-i\mo\cdot z \hv}  \varphi_{0} \right)(x)=  \varphi_{0}(x-z\hv).
\label{3.26}
\ene
The term $ e^{ -i \frac{z}{v} H_0}$ gives raise to the quantum-mechanical spreading of the wave packet. For high velocities this  term is one
order of magnitude smaller than the classical translation, and if we neglect it we get that,
\beq
(e^{ -i \frac{z}{v} H_0}\, \varphi_{\v})(x) \approx  e^{i\frac{mzv}{2}}\,   \varphi_{\v}(x-z\hv), \,\, \hbox{\rm for large}\,\, v.
\label{3.27}
\ene
We see that, in this approximation, for high velocities our asymptotic state evolves along the classical trajectory, modulo the global phase factor
 $e^{i\frac{mzv}{2}}$ that plays no
role. The key issue is that the support of our incoming wave packet
remains in $\Lambda_{\v}$ for all distances, or for all times, and
in consequence it has no interaction with $K$. We can expect that
for high velocities  the exact solution, $\psi_{\v}$ (\ref{3.21}),
to the interacting Schr\"odinger equation (\ref{2.14})  is close to
the incoming wave packet $\psi_{\v,0}$ and that, in consequence, it
also has negligible interaction with $K$, provided, of course,  that
the support of $\varphi_0$ is contained in $\Lambda_{\v}$. Below we
give a rigorous ground for this heuristic picture proving that in
the leading order $\psi_{\v}$ is not influenced by $K$ and  that it
only contains information on the potential $A$.

We define,
\beq
W_{\pm,\v}(A,V):= e^{-im\v\cdot x}\, W_{\pm}(A,V)\, e^{im\v\cdot x} =  \hbox{\rm s-}\lim_{z \rightarrow \pm \infty} e^{iz H_2(A,V)}\, J\, e^{-iz H_1}.
\label{3.28}
\ene
\begin{lemma}\label{lemm-3.3}
Let $\Lambda_0$ be a compact subset of $\Lambda_{\hv}, \v \in \ere^3\setminus 0$. Then, for all $A \in \p2$ and all $\chi \in C^{\infty}(\ere^3)$
that satisfies $\chi(x)= 0$ for  $x$ in a neighborhood of $K$, $\chi(x)=1,\, \hbox{\rm for}\, x \in \{ x: x=y+\tau \hv, y\in \Lambda_0, \tau \in \ere \} \cup
\{ x: |x| \geq R \}$ with $R$ such that $ K \subset B_R$, there is a constant $C$ such that,
\beq
\left\|e^{-i \frac{z}{v}\, H(A,V)}\, W_{\pm}(A,V) \,\varphi_{\v} - \chi e^{-i L_{A,\hv}(\pm \infty)}\,e^{-i \frac{z}{v}\,H_0} \varphi_\v \right\|
\leq  \frac{C}{v}\, (1+ (1 \mp\hbox{\rm sign}(z))|z|)\, \|\varphi\|_{{\mathcal H}_2(\ere^3)},
\label{3.29}
\ene
for all $z \in \ere$ and all $\varphi \in {\mathcal H}_2(\ere^3)$ with support contained in $\Lambda_0$.
\end{lemma}

\noindent {\it Proof:} By (\ref{3.16}) it is enough to prove the lemma for $\tilde{\varphi}$. We first give the proof for a potential $A \in \p2$ that
satisfies
\beq
|A(x)|+|\hbox{\rm div} A(x)|\leq C (1+|x|)^{-\beta_1}, \quad \beta_1 >1,
\label{3.30}
\ene
for example, for the Coulomb potential.

By the intertwining relations (\ref{2.20})
\beq
\begin{array}{c}
e^{-i \frac{z}{v}\, H(A,V)}\, W_{\pm}(A,V) \,\tvf_{\v} - \chi e^{-i L_{A,\hv}(\pm \infty)}\,e^{-i \frac{z}{v}\,H_0} \tvf_\v=\\\\
e^{im\v\cdot x}
 \hbox{\rm s-}\, \lim_{t \rightarrow \pm \infty}\left[ e^{i t H_2}\chi (x)e^{-i t H_1}- \chi(x)e^{-iL_{A,\hv}
 (t)}
 \right]\, e^{-i z H_1}
 \tvf .
\label{3.31}
\end{array}
\ene
Denote,
\beq
\begin{array}{c}
P(t,\tau, z):= e^{i(\tau-z) H_2} i \left[H_{2}e^{-i L_{A, \hv}(t-(\tau-z))}\chi(x)-
\right. \\\\
\left.
e^{-i L_{A,\hv}(t-(\tau-z))} \chi(x)
\left(H_1-\hv\cdot A(x+(t-(\tau -z))\hv)\right)\right] e^{-i\tau H_1}\tvf.
\label{3.32}
\end{array}
\ene
Then, by Duhamel's formula - see equation (5.26) of \cite{bw} and \cite{w1}-
\beq
\left[ e^{it H_2}\chi (x)e^{-it H_1}- \chi(x)e^{-iL_{A,\hv}
 \left(t\right)}
 \right]
 \tvf = \int_z^{t+z}\, d \tau\, P(t,\tau,z).
 \label{3.33}
 \ene

We have that (see equations (5.29-5.32) of \cite{bw}  and \cite{w1}),
\beq
P(t,\tau,z) = T_1+T_2+T_3
\label{3.34},
\ene
where
\beq
T_1:= \frac{1}{v} e^{i (\tau-z) H_2}i e^{-iL_{A ,\hv}(x,t-(\tau-z))}\, \left( \Xi_b(x,t-(\tau-z))+ \chi V(x)\right)e^{-i\tau H_1}\tvf ,
\label{3.35}
\ene
\beq
\begin{array}{c}
T_2:= \frac{1}{2mv}e^{i (\tau-z) H_2}i e^{-i L_{A , \hv}(x,t-(\tau-z))}\left\{-(\Delta \chi)+2(\mo \chi)\cdot \mo
-2 b(x, t-(\tau-z)) \cdot (\mo \chi)\right\} e^{-i\tau H_1}\tvf,
\label{3.36}
\end{array}
\ene

\beq
T_3:=e^{i (\tau-z) H_2}i e^{-i L_{A,\hv}(x,t-(\tau-z))} \left[ (\mo \chi)\cdot \hv  \right]
e^{-i\tau H_1}\tvf.
\label{3.37}
\ene
Note that,
\beq
\begin{array}{c}
\left|\eta(x,t-(\tau-z))\, F(|x-\tau \hv|\leq |\tau/4|)\right| \leq C (1+|\tau|)^{-\mu+1},\\\\
 \hbox{if}\,\,  t+z \geq 0 \,\hbox{\rm and}\,
\tau \in [0,t+z ]\, \hbox{or if}\,\, t+z \leq 0\, \hbox{\rm and}\, \tau \in [t+z,0].
\label{3.38}
\end{array}
\ene
Furthermore, since $\nabla\cdot (\hv \times B)= - \hv \cdot \hbox{\rm curl}\, B$,
\beq
\begin{array}{c}
\left|\mo\cdot \eta(x,t-(\tau-z)) F(|x-\tau \hv|\leq |\tau/4|)    \right| \leq C (1+\tau)^{-\mu+1},\\\\
\hbox{if}\,  t+z \geq 0 \, \hbox{\rm and}\,
\tau \in [0,t+z ]\, \hbox{or if }\, t+z \leq 0\, \hbox{\rm and}\, \tau \in [t+z,0].
\end{array}
\label{3.39}
\ene

We give the proof for $W_+(A,V)$. The case of $W_-(A,V)$ follows in the same way. Since we have to take the limit $t \rightarrow \infty$ in
(\ref{3.31}), we can assume that $ t > 2|z|$.
Let us estimate
$$
\| \int_z^{t+z} \, T_1 d\tau \|.
$$
We consider first the terms in $\Xi_{b}$ that do not contain $A$. for example the term,
$$
I_1:= \frac{-1}{m v} \int_z^{t+z}\, d \tau\, e^{i (\tau-z) H_2}i e^{-iL_{A ,\hv}(x,t-(\tau-z))}\, \chi(x)\,\eta(x,t-(z-\tau))\cdot \mo  e^{-i\tau H_1}\tvf.
$$
We have that,
$$
\begin{array}{c}
\|I_1\| \leq  \frac{1}{m v}\, \int_z^{0}\, d \tau \, \|\eta(x,t-(z-\tau))\cdot \mo  e^{-i\tau H_1}\tvf\|+\\\\
\frac{1}{m v}\, \int_0^{t+z}\, d \tau \, \|\eta(x,t-(z-\tau))\cdot \mo  e^{-i\tau H_1}\tvf\| \leq \frac{C}{v} (1+ (1-\hbox{\rm sign}(z)) |z|)\,
\|\tvf\|_{\mathcal H^1(\ere^3)},
\end{array}
$$
where we used (\ref{3.13}) and (\ref{3.38}). Let us now estimate a term in $\Xi_b$ that contains $A$. for example,
$$
I_2:= \frac{-1}{m v}  \int_z^{t+z}\, d \tau\, e^{i (\tau-z) H_2}i e^{-iL_{A ,\hv}(x+t-(\tau-z))}\, \chi(x)\,A(x+(t-(z-\tau))\hv)\cdot \mo
 e^{-i\tau H_1}
\tvf.
$$
Since, $z \leq \tau \leq t+z$ and $t \geq 2|z|$ we have that $|\tau| \leq t+z$. Then, for $ |x-\tau \hv| \leq |\tau|/4$ we have that,
$|x+(t-(\tau-z))\hv)| \geq |t+z|-|\tau|/4\geq 3 |\tau|/4$. Then by (\ref{3.13}, \ref{3.30})
$$
\|I_2\|  \leq  \frac{C}{v}\|\tvf\|_{\mathcal H^1(\ere^3)}.
$$
The remaining terms in $T_1$ are estimated in the same way, using (\ref{3.11}) in the term containing $\chi V$. in this way we prove that,
\beq
\| \int_z^{t+z}T_1\| \leq \frac{C}{v} (1+ (1-\hbox{\rm sign}(z)) |z|)\,
\|\tvf\|_{\mathcal H^2(\ere^3)}.
\label{3.40}
\ene
In the same way we prove that,
\beq
\| \int_z^{t+z}T_2\| \leq \frac{C}{v} \|\tvf\|_{\mathcal H^1(\ere^3)}.
\label{3.41}
\ene
Moreover, by equation (5.37) of \cite{bw} (see also the proof of Lemma 2.4 of \cite{w1}),
\beq
\|\int_z^{t+z} \, T_3(\tau)\| \leq \frac{C}{v}
\|\phi\|_{\mathcal H_2(\ere^3)}.
\label{3.42}
\ene
Note that it is in the proof of (\ref{3.42}) that the condition
$\chi(x)=1,\, \hbox{\rm for}\, x \in \{ x: x=y+\tau \hv, y\in \Lambda_0, \tau \in \ere \}$ is used.
Equation (\ref{3.29}) follows from (\ref{3.34}-\ref{3.37}) and (\ref{3.40}-\ref{3.42}).

Let us now consider the case of $A \in \p2$. We take  $\tilde{A}\in \p2$ that satisfies (\ref{3.30}) and has the same fluxes as $A$.
Let $\lambda$ be as in (\ref{2.11}). We give the proof for $W_+(A,V)$. The case of $W_-(A,V)$ is similar. By the gauge transformation formula
(\ref{2.21}),
\beq
\begin{array}{c}
\left\|e^{-i \frac{z}{v}\, H(A,V)}\, W_{+}(A,V) \,\varphi_{\v} - \chi e^{-i L_{A,\hv}( \infty)}\,e^{-i \frac{z}{v}\,H_0} \varphi_\v\right\|=\\\\
\left\|e^{-i\lambda(x) }e^{-i \frac{z}{v}\, H(\tilde{A},V)}\, W_{+}(\tilde{A},V) \, e^{i\lambda_\infty(\mo)} \,\varphi_{\v} -
\chi e^{-i L_{\tilde{A},\hv}( \infty)}\, e^{i\lambda_\infty(\hv)}\, e^{-i\lambda(x)}\,
e^{-i \frac{z}{v}\,H_0} \varphi_\v\right\| \leq \\\\
 \frac{C}{v}\, (1+ (1 -\hbox{\rm sign}(z))|z|)\, \|\varphi\|_{{\mathcal H}_2(\ere^3)}+ C \left\| \left(e^{i\lambda_\infty(\mo)}- e^{i\lambda_\infty(\hv)}
 \right) \varphi_\v\right\|_{L^2(\ere^3)}.
 \label{3.43}
 \end{array}
\ene
But, by (\ref{2.12}), (\ref{3.7}) and since $\lambda_\infty$ is homogenous of degree zero,
\beq
\left\| \left(e^{i\lambda_\infty(\mo)}- e^{i\lambda_\infty(\hv)}
 \right) \varphi_\v\right\|_{L^2(\ere^3)} \leq \frac{C}{v}\, _{\mathcal H_1(\ere^3)}.
\label{3.44}
\ene
Equation (\ref{3.29}) follows from (\ref{3.43}, \ref{3.44}).

\begin{lemma} \label{lemm-3.4}
Suppose that $ A \in \p2$. Then, there is a constant $C$ such that,
\beq
\left\|  \left( e^{-i L_{A,\hv}(\pm\infty )} -1 \right)\, e^{- i \frac{z}{v} H_0 } \varphi_\v \right\|_{L^2(\ere^3)}
 \leq C\left((1+|z|)^{-\beta_l+1}+ \frac{1}{v}\right)
\,\|\varphi\|_{\mathcal H_2(\ere^3)},\quad \hbox{for}\, \pm z >0,
\label{3.45}
\ene
and all $\varphi \in \mathcal H_2(\ere^3)$.
\end{lemma}
\noindent {\it Proof:}
By (\ref{3.16}) it is enough to prove the lemma for $\tvf$. We give the proof in the $+$ case. The $-$ case follows in the same way. By
(\ref{3.7}, \ref{3.9}) we have that,
\beq
\left\|  \left( e^{-i L_{A,\hv}(\infty )} -1 \right)\, e^{- i \frac{z}{v} H_0 } \tvf_\v \right\|_{L^2(\ere^3)}\leq
\left\| \left( \int_0^\infty (A\cdot\hv )(x+\tau \hv)\, d\tau  \right)\, e^{- i z H_1} \tvf \right\|_{L^2(\ere^3)}.
\label{3.46}
\ene
Furthermore, denoting $ x=x_{\parallel}\hv + x_
\perp$, where $x_{\parallel}$
is the component of $x$ parallel to $\hv$ and $x_\perp$ is the component of $x$ perpendicular to $\hv$, it follows from (\ref{3.6}) that,
\beq
\begin{array}{l}
\left\| F(|x- z \hv|< |z|/4)\left(\int_0^\infty (A\cdot\hv )(x+\tau \hv) \, d\tau \right) \right\|_{{\mathcal B}(\ere^3)} =
\left\| F(|x- z \hv |< |z|/4) e^{i \mo\cdot x_{\perp}}
\left(\int_{x_{\parallel}}^\infty (A\cdot\hv )(\tau \hv )\, d\tau\right)\, e^{-i \mo\cdot x_{\perp}} \right\|_{{\mathcal B}(\ere^3)} \\\\
= \left\| F(|x_{\parallel}-z|< |z|/4)
\left(\int_{x_{\parallel}}^\infty (A\cdot\hv)(\tau \hv)\, d\tau\right) \right\|_{{\mathcal B}(\ere^3)} \leq
\int_{3z/4}^\infty \, \left|(A\cdot\hv)(\tau \hv)\right|\, d\tau
\leq  C\, (1+z)^{-\beta_l+1}.
\end{array}
\label{3.47}
\ene
The lemma follows from (\ref{3.46}, \ref{3.47}) and Lemma \ref{lemm-3.2}.
\begin{lemma}\label{lemm-3.5}
Let $\Lambda_0$ be a compact subset of $\Lambda_{\hv}, \v \in \ere^3\setminus 0$. Then, for all $A \in \p2$ , there is a constant $C$ such that,
\beq
\left\|e^{-i \frac{z}{v}\, H(A,V)}\, W_{\pm}(A,V) \,\varphi_{\v} - e^{-i \frac{z}{v}\,H_0} \varphi_\v \right\|
\leq   C\left( \frac{1}{v} + (1+|z|)^{-\beta_l+1}\right) \|\varphi\|_{{\mathcal H}_2(\ere^3)},\quad \hbox{for}\, \pm z >0,
\label{3.48}
\ene
and all $\varphi \in {\mathcal H}_2(\ere^3)$ with support contained in $\Lambda_0$.
\end{lemma}

\noindent {\it Proof:} The Lemma follows from  Lemmata \ref{lemm-3.3}, \ref{lemm-3.4}, (\ref{3.16}) and since by Lemma \ref{lemm-3.2}
\beq
\left\|(1-\chi) e^{-i\frac{z}{v}H_0} \tvf_\v\right\| \leq C_l (1+|z|)^{-l}\, \|\tvf\|_{L^2(\ere^3)}, \quad l=1,2,\cdots.
\label{3.49}
\ene

\begin{lemma}\label{lemm-3.6}
Let $\Lambda_0$ be a compact subset of $\Lambda_{\hv}, \v \in \ere^3\setminus 0$. Then, for all $A \in \p2$ with \,
$\hbox{\rm div} A \in L^2\left(\overline{\Lambda}\right)$ \,there is a constant $C$ such that,
\beq
\left\|e^{-i \frac{z}{v}\, H(A,V)}\, W_{-}(A,V) \,\varphi_{\v} - e^{-i \frac{z}{v}\,H_0} \, e^{i\int_{-\infty}^\infty A\cdot\hv(x+\tau \hv)\, d\tau }
\,\varphi_\v \right\|
\leq   C\left( \frac{1}{v} + (1+z)^{-\beta_l+1}\right) \|\varphi\|_{{\mathcal H}_2(\ere^3)},\quad \hbox{for}\,  z \geq 0,
\label{3.50}
\ene
and all $\varphi \in {\mathcal H}_2(\ere^3)$ with support contained in $\Lambda_0$.
\end{lemma}

\noindent{\it Proof:} First  note that

$$
\int_{-\infty}^\infty A\cdot\hv(x+\tau \hv)\, d\tau = L_{A,\hv}(\infty)- L_{A,\hv}(-\infty).
 $$

By equations (5.19) and (5.42) of \cite{bw},
\beq \label{3.51}
\left\| W_{-}(A,V) \,\varphi_{\v} -  W_{+}(A,V) \,e^{i L_{A,\hv}(\infty)-i L_{A,\hv}(-\infty)}
\,\varphi_\v \right\| \leq \frac{C}{v}\, \left \|  \varphi_\v  \right\|_{{\mathcal H}_2(\ere^3)}.
\ene
Then,
\beq\label{3.52}
\begin{array}{l}
\left\|e^{-i \frac{z}{v}\, H(A,V)}\, W_{-}(A,V) \,\varphi_{\v} - e^{-i \frac{z}{v}\,H_0} \, e^{i\int_{-\infty}^\infty A\cdot\hv(x+\tau \hv)\, d\tau }
\,\varphi_\v \right\|
\leq   C \frac{1}{v} \|\varphi\|_{{\mathcal H}_2(\ere^3)}+\\\\
 \left\|e^{-i \frac{z}{v}\, H(A,V)}\, W_{+}(A,V) \, e^{i L_{A,\hv}(\infty)-i L_{A,\hv}(-\infty)} \,\varphi_{\v}
- e^{-i \frac{z}{v}\,H_0} \,  e^{i L_{A,\hv}(\infty)-i L_{A,\hv}(-\infty)}
\,\varphi_\v \right\| \leq   C \frac{1}{v} \|\varphi\|_{{\mathcal H}_2(\ere^3)}+\\\\
C \left( \frac{1}{v}+ (1+z)^{-\beta_l+1}\right) \|\varphi\|_{{\mathcal H}_2(\ere^3)},\quad \hbox{for}\, z >0,
\end{array}
\ene
were we used Lemma \ref{lemm-3.5} and equation (5.42) of \cite{bw}.

\begin{lemma}\label{lemm-3.7}
For all $A \in \p2$ with \,
$\hbox{\rm div} A \in L^2\left(\overline{\Lambda}\right) $there is a constant $C$ such that, $\forall z \in \ere$,
\beq
\left\| e^{i\int_{-\infty}^\infty A\cdot\hv(x+\tau \hv)\, d\tau }\, e^{-i \frac{z}{v}\,H_0} \varphi_\v - e^{-i \frac{z}{v}\,H_0}\,
e^{i\int_{-\infty}^\infty A\cdot\hv(x+\tau \hv)\, d\tau }\, \varphi_\v
\right\| \leq C  \frac{|z|}{v}  \|\varphi\|_{\mathcal H_2(\ere^3)},
\label{3.53}
\ene
for all $ \varphi \in \mathcal H_2(\ere^3)$.
\end{lemma}

\noindent{\it Proof:}
By (\ref{3.9})
\beq
\begin{array}{l}
N:=\left\| e^{i\int_{-\infty}^\infty A\cdot\hv(x+\tau \hv)\, d\tau }\, e^{-i \frac{z}{v}\,H_0} \varphi_\v - e^{-i \frac{z}{v}\,H_0}\,
e^{i\int_{-\infty}^\infty A\cdot\hv(x+\tau \hv)\, d\tau }\, \varphi_\v
\right\|=\\\\
\left\| e^{i\int_{-\infty}^\infty A\cdot\hv(x+\tau \hv)\, d\tau }\, e^{-i z\,H_1} \varphi - e^{-i z\,H_1}\,
e^{i\int_{-\infty}^\infty A\cdot\hv(x+\tau \hv)\, d\tau }\, \varphi \right\|.
\end{array}
\label{3.54}
\ene
Moreover by (\ref{3.8}),
\beq
\left\|e^{i\int_{-\infty}^\infty A\cdot\hv(x+\tau \hv)\, d\tau }\,\left( e^{-i z\,H_1}- e^{-i( z\mo\cdot \hv+ mv z/2)}\right)\varphi\right\|
\leq \left\| \frac{z}{v}\, H_0 \varphi\right\| \leq C \frac{|z|}{v}\, \|\varphi\|_{\mathcal H_2(\ere^3)}.
\label{3.55}
\ene
Furthermore, by (\ref{3.6})
\beq
e^{i\int_{-\infty}^\infty A\cdot\hv(x+\tau \hv)\, d\tau }\, e^{-i(z \mo\cdot \hv+ mv z/2)}\varphi=
e^{-i( z \mo\cdot \hv+ mv z/2)}\, e^{i\int_{-\infty}^\infty A\cdot\hv(x+\tau \hv)\, d\tau }\,\varphi.
\label{3.56}
\ene
Then by (\ref{3.54}, \ref{3.55}, \ref{3.56}),
\beq
\begin{array}{l}
N \leq \left\|e^{-i( z\mo\cdot \hv+ mv z/2)}\, e^{i\int_{-\infty}^\infty A\cdot\hv(x+\tau \hv)\, d\tau }\, \varphi-
e^{-i z\,H_1}\,
e^{i\int_{-\infty}^\infty A\cdot\hv(x+\tau \hv)\, d\tau }\, \varphi \right\|+ C \frac{|z|}{v}\, \|\varphi\|_{\mathcal H_2(\ere^3)} \leq\\\\
 C \frac{|z|}{v}\, \|\varphi\|_{\mathcal H_2(\ere^3)}+ \left\|
 \left( e^{-i z\,H_1}- e^{-i( z\mo\cdot \hv+ mv z/2)}\right) e^{i\int_{-\infty}^\infty A\cdot\hv(x+\tau \hv)\, d\tau }\   \varphi\right\|
\leq C \frac{|z|}{v}\, \|\varphi\|_{\mathcal H_2(\ere^3)},
\end{array}
\label{3.57}
\ene
where we used equation (5.42) of \cite{bw}.
\begin{lemma}\label{lemm-3.8}
Let $\Lambda_0$ be a compact subset of $\Lambda_{\hv}, \v \in \ere^3\setminus 0$. Then, for all $A \in \p2$
with \, $\hbox{\rm div} A \in L^2\left(\overline{\Lambda}\right)$there is a constant $C$ such that $ \forall\,  z \geq Z \geq 0,$
\beq
\left\|e^{-i \frac{z}{v}\, H(A,V)}\, W_{-}(A,V) \,\varphi_{\v} - e^{-i \frac{z-Z}{v}\,H_0} \, e^{-i L_{A,\hv}(-\infty)}
\,  e^{-i \frac{Z}{v}\,H_0}\, \varphi_\v \right\|
\leq   C\left( \frac{1}{v} + (1+Z)^{-\beta_l+1}+ \frac{Z}{v}\right) \|\varphi\|_{{\mathcal H}_2(\ere^3)},
\label{3.58}
\ene
and all $\varphi \in {\mathcal H}_2(\ere^3)$ with support contained in $\Lambda_0$.
\end{lemma}

\noindent{\it Proof:} The lemma follows from Lemmata \ref{lemm-3.4}, \ref{lemm-3.6} and \ref{lemm-3.7} and equation (5.42) of \cite{bw}.

\bull

We summarize the results that we have obtained in the following theorem.

\begin{theorem}\label{theor-3.9}
Let $\Lambda_0$ be a compact subset of $\Lambda_{\hv}, \v \in \ere^3\setminus 0$. Then, for all $A \in \p2$
there is a constant $C$ such that the following estimates hold for all
$\varphi \in {\mathcal H}_2(\ere^3)$ with support contained in $\Lambda_0$.
\begin{enumerate}
\item
For all $ Z \geq 0$ and all $ z \leq Z$,
\beq
\left\|e^{-i \frac{z}{v}\, H(A,V)}\, W_{-}(A,V) \,\varphi_{\v} -  e^{-i L_{A,\hv}(- \infty)}\,e^{-i \frac{z}{v}\,H_0} \varphi_\v \right\|
\leq  \frac{C}{v}\, (1+ Z)\, \|\varphi\|_{{\mathcal H}_2(\ere^3)}.
\label{3.59}
\ene
If furthermore, \, $\hbox{\rm div} A \in L^2\left(\overline{\Lambda}\right)$,
\item
For all $ Z \geq 0$ and all $ z \geq Z$,
\beq
\begin{array}{l}
\left\|e^{-i \frac{z}{v}\, H(A,V)}\, W_{-}(A,V) \,\varphi_{\v} - e^{-i \frac{z-Z}{v}\,H_0} \, e^{-i L_{A,\hv}(-\infty) }\,e^{-i \frac{Z}{v}\,H_0}
\,\varphi_\v \right\|\\\\
\leq   C\left( \frac{1}{v} + (1+Z)^{-\beta_l+1}+ \frac{Z}{v}\right) \|\varphi\|_{{\mathcal H}_2(\ere^3)}.
\end{array}
\label{3.60}
\ene
\item
For all $z \geq  0$,
\beq
\left\|e^{-i \frac{z}{v}\, H(A,V)}\, W_{-}(A,V) \,\varphi_{\v} - e^{-i \frac{z}{v}\,H_0} \, e^{i\int_{-\infty}^\infty A\cdot\hv(x+\tau \hv)\, d\tau }
\,\varphi_\v \right\|
\leq   C\left( \frac{1}{v} + (1+z)^{-\beta_l+1}\right) \|\varphi\|_{{\mathcal H}_2(\ere^3)}.
\label{3.61}
\ene
\end{enumerate}
\end{theorem}
\noindent {\it Proof:} The theorem follows from   equation (\ref{3.49}) and Lemmata \ref{3.3}, \ref{3.6} and \ref{3.8}.
\begin{theorem}\label{theor-3.10}
Let $\Lambda_0$ be a compact subset of $\Lambda_{\hv}, \v \in \ere^3\setminus 0$. Then, for all $A \in \p2$  there is a constant $C$ such that the following estimates hold for all
$\varphi \in {\mathcal H}_2(\ere^3)$ with support contained in $\Lambda_0$.
\begin{enumerate}
\item
For all  $ z \leq  v^{1/\beta_l}$,
\beq
\left\|e^{-i \frac{z}{v}\, H(A,V)}\, W_{-}(A,V) \,\varphi_{\v} -  e^{-i L_{A,\hv}(- \infty)}\,e^{-i \frac{z}{v}\,H_0} \varphi_\v \right\|
\leq  \frac{C}{v^{1-1/\beta_l}}\, \|\varphi\|_{{\mathcal H}_2(\ere^3)}.
\label{3.62}
\ene
If furthermore, \, $\hbox{\rm div} A \in L^2\left(\overline{\Lambda}\right)$,
\item
For all $z \geq v^{1/\beta_l}$,
\beq
\begin{array}{l}
\left\|e^{-i \frac{z}{v}\, H(A,V)}\, W_{-}(A,V) \,\varphi_{\v} - e^{-i \frac{z-v^{1/\beta_l}}{v}\,H_0} \, e^{-i L_{A,\hv}(-\infty) }
\,e^{-i \frac{v^{1/\beta_l}}{v}\,H_0}
\,\varphi_\v \right\|\\\\
\leq   \frac{C}{v^{1-1/\beta_l}}\, \|\varphi\|_{{\mathcal H}_2(\ere^3)}.
\end{array}
\label{3.63}
\ene
\item
For all $z \geq v^{1/\beta_l}$,
\beq
\left\|e^{-i \frac{z}{v}\, H(A,V)}\, W_{-}(A,V) \,\varphi_{\v} - e^{-i \frac{z}{v}\,H_0} \, e^{i\int_{-\infty}^\infty A\cdot\hv(x+\tau \hv)\, d\tau }
\,\varphi_\v \right\|
\leq    \frac{C}{v^{1-1/\beta_l}}\,  \|\varphi\|_{{\mathcal H}_2(\ere^3)}.
\label{3.64}
\ene
\end{enumerate}
\end{theorem}
\noindent {\it Proof:} In Theorem \ref{theor-3.9} we take $Z= v^\rho, 0 < \rho <1$. The error terms are of the form, $ 1/v, 1/ v^{\rho (\beta_l-1)}$
and $ 1/ v^{1-\rho}$. As for $ v \geq 1$ the error $1/v$ is smaller than $ 1/ v^{1-\rho}$ we only have to consider $1/ v^{\rho (\beta_l-1)}$
and $ 1/ v^{1-\rho}$. Looking to these errors as a function of $\rho$ we see that the point where the smallest exponent is bigger is the point of intersection of the lines
$1-\rho$ and $ \rho(\beta_l-1)$, i.e., $ 1-\rho= \rho(\beta_l-1)$. Hence we take, $\rho= 1/ \beta_l$. The theorem follows from Theorem \ref{theor-3.9}.

\subsection{Physical Interpretation}

In Theorems \ref{theor-3.9} and \ref{theor-3.10} we give the leading order for high-velocity of the solution to the Schr\"odinger equation. In
equation (\ref{3.59}) we give the leading order when the electron is incoming and interacting.  We see that as the solution propagates towards  the
 magnet, and it crosses it, it picks up a phase. In equations (\ref{3.60}, \ref{3.61}) we give two different expressions for the leading order when the
 electron is outgoing, i.e. after it leaves the magnet. The distance $Z$ separates the incoming and interacting region from the outgoing one.
 In equation (\ref{3.60}) we see that the leading order for the outgoing electron at distance $z$ consists of the incoming and interacting leading order
 taken as the initial data at distance $Z$ followed by the free evolution during distance $z-Z$. Finally, in equation (\ref{3.61}) we give another
  representation of the leading order of the
 outgoing electron. Recall that the Cauchy data of the outgoing solution is given $ S\varphi_\v$, with $S$ the scattering operator.
 Furthermore (see Theorem 5.7 of \cite{bw}), up to an error of order $1/v$,  $S  \varphi_\v= e^{i\int_{-\infty}^\infty A\cdot\hv(x+\tau \hv)\, d\tau }
 \, \varphi_\v$. Then, equation (\ref{3.61}) expresses the leading order when the electron
is outgoing as the free evolution applied to the Cauchy data of the outgoing solution. Note that scattering theory  and Theorem 5.7 of \cite{bw} tell us
that, up to an error of order $1/ v$, the interacting solution
tends to $e^{-i\frac{z}{v} H_0}\, e^{i\int_{-\infty}^\infty A\cdot\hv(x+\tau \hv)\, d\tau }\, \varphi_\v$ at $t \rightarrow \infty$. Equation
 (\ref{3.61})
is more precise. It actually gives us an estimate of the error bound for large distances.

Note that the leading orders for the outgoing electron given in equations (\ref{3.60}, \ref{3.61}) are close to each other for high velocity.
It follows from Lemmata \ref{lemm-3.4} and
\ref{lemm-3.7} that for $z\in \ere,   Z \geq 0$,
\beq
\begin{array}{l}
\left\|  e^{-i \frac{z-Z}{v}\,H_0} \, e^{-i L_{A,\hv}(-\infty) }\,e^{-i \frac{Z}{v}\,H_0}
\,\varphi_\v - e^{-i \frac{z}{v}\,H_0} \, e^{i\int_{-\infty}^\infty A\cdot\hv(x+\tau \hv)\, d\tau }
\,\varphi_\v \right\| \\\\ \leq   C\left( \frac{1}{v} + (1+Z)^{-\beta_l+1}+ \frac{Z}{v}\right) \|\varphi\|_{{\mathcal H}_2(\ere^3)}.
\end{array}
\label{3.65}
\ene
In equations (\ref{3.62}, \ref{3.63}, \ref{3.64}) we optimize the error bounds taking the transition distance as
$ Z= v^{1/\beta_l}$ and we obtain high-velocity estimates that are uniform, respectively, for $ z \leq  v^{1/\beta_l} $, and $z \geq  v^{1/\beta_l} $.
Furthermore, taking $Z= v^{1/\beta_l} $  in (\ref{3.65}) we obtain
\beq
\begin{array}{l}
\left\|  e^{\ds -i \frac{z-v^{1/\beta_l}}{v}\,H_0} \, e^{-i L_{A,\hv}(-\infty) }\,e^{\ds -i \frac{v^{1/\beta_l}}{v}\,H_0}
\,\varphi_\v - e^{-i \frac{z}{v}\,H_0} \, e^{i\int_{-\infty}^\infty A\cdot\hv(x+\tau \hv)\, d\tau }
\,\varphi_\v \right\| \\\\
 \leq    \frac{C}{v^{1-1/\beta_l}}\, \|\varphi\|_{{\mathcal H}_2(\ere^3)}, \quad z \in \ere.
 \end{array}
\label{3.66}
\ene
In the transition region around $Z$ the different expressions that we have obtained for the leading order are close to each other, as we show
 in the next sub-subsection.

\subsubsection{The Transition Region}
We estimate the difference between the leading orders in Theorems \ref{theor-3.9} and \ref{theor-3.10} in the transition region
$z \in [ Z/L,  Z L ], Z,  L>1 $.

It follows from Lemmata \ref{lemm-3.4}, \ref{lemm-3.7} and from equation ( 5.42) of \cite{bw} that for $ z \in[Z/L,Z L]$,
\beq
\left\| e^{-i L_{A,\hv}(- \infty)}\,e^{-i \frac{z}{v}\,H_0} \varphi_\v -   e^{-i \frac{z}{v}\,H_0} \, e^{i\int_{-\infty}^\infty A\cdot\hv(x+\tau \hv)\, d\tau }
\,\varphi_\v \right\| \leq C \left((1+Z/L)^{-\beta_l+1}+ \frac{1+ Z L}{v}\right)\,  \|\varphi\|_{\mathcal H_2(\ere^3)}.
\label{3.67}
\ene
In the same way we prove that that for $ z \in[Z/L,Z L], v >1$,
\beq
\left\|  e^{-i L_{A,\hv}(- \infty)}\,e^{-i \frac{z}{v}\,H_0} \varphi_\v - e^{-i \frac{z-Z}{v}\,H_0} \, e^{-i L_{A,\hv}(-\infty) }\,e^{-i \frac{Z}{v}\,H_0}
\,\varphi_\v \right\|  \leq C \left( (1+Z/L)^{-\beta_l+1}+ \frac{1+ Z L}{v}\right)\,\|\varphi\|_{\mathcal H_2(\ere^3)}.
\label{3.68}
\ene
Taking as in Theorem \ref{theor-3.10},   $Z=  v^{1/\beta_l}$, we obtain that for $ z \in [\frac{ v^{1/\beta_l}}{ L}, L  v^{1/\beta_l}  ]$,
\beq
\begin{array}{l}
\left\| e^{-i L_{A,\hv}(- \infty)}\,e^{-i \frac{z}{v}\,H_0} \varphi_\v -   e^{-i \frac{z}{v}\,H_0} \, e^{i\int_{-\infty}^\infty A\cdot\hv(x+\tau \hv)\,
 d\tau }
\,\varphi_\v \right\| \\\\
 \leq C \left( L^{\beta_l-1}+ 1+L \right)\, \frac{1}{v^{1-1/\beta_l}}\,
\|\varphi\|_{\mathcal H_2(\ere^3)},
\end{array}
\label{3.69}
\ene
\beq
\begin{array}{l}
\left\|  e^{-i L_{A,\hv}(- \infty)}\,e^{-i \frac{z}{v}\,H_0} \varphi_\v - e^{-i \frac{z-Z}{v}\,H_0} \, e^{-i L_{A,\hv}(-\infty) }
\,e^{-i \frac{Z}{v}\,H_0}
\,\varphi_\v \right\|  \\\\
\leq C \left( L^{\beta_l-1}+ 1+L\right)\, \frac{1}{v^{1-1/\beta_l}}\,
\|\varphi\|_{\mathcal H_2(\ere^3)}.
\end{array}
\label{3.70}
\ene
\subsection{Final Formulae}
Summing up, we have proven in Theorems \ref{theor-3.9} and \ref{theor-3.10} that  the leading order for high velocity of the exact solution to the
Schr\"odinger equation, $\psi_{\v}= e^{-it H(A,V)}\, W_-(A,V)\, \varphi_{\v}$, that behaves as, $\psi_{\v,0}:= e^{-itH_0} \, \varphi_\v$,
 when $t \rightarrow -\infty$, is given by the following approximate solution to the Schr\"odinger equation,
\beq
\psi_{\v,App}(x,z):=\left\{
\begin{array}{l}
e^{-i L_{A,\hv}(- \infty)}\,e^{-i \frac{z}{v}\,H_0} \varphi_\v,\quad   z= vt \leq Z \geq 0,\\\\
e^{-i \frac{z-Z}{v}\,H_0} \, e^{-i L_{A,\hv}(-\infty) }\,e^{-i \frac{Z}{v}\,H_0}
\,\varphi_\v, \quad z=vt \geq Z,
\end{array}\right.
\label{3.71}
\ene
and, equivalently, by the approximate solution,
\beq
\phi_{\v,App}(x,z):=\left\{
\begin{array}{l}
e^{-i L_{A,\hv}(- \infty)}\,e^{-i \frac{z}{v}\,H_0} \varphi_\v,\quad   z= v t \leq Z \geq 0,\\\\
 e^{-i \frac{z}{v}\,H_0} \, e^{i\int_{-\infty}^\infty A\cdot\hv(x+\tau \hv)\, d\tau }
\,\varphi_\v, \quad z=vt \geq Z.
\end{array}\right.
\label{3.72}
\ene

\section{The Aharonov-Bohm Effect}\sss
We will consider now the case where the magnetic field, $B$, outside $K$ is zero but with a non-trivial magnetic flux, $\Phi$, inside $K$. For the moment we also
suppose that the electric potential, $V$, outside $K$ is zero, but this actually is not essential as the electric potential gives rise to a lower order
effect for high velocity. This situation corresponds to the Aharonov-Bohm effect \cite{ab} and in particular to the experiments of Tonomura et al.
\cite{to3}, \cite{to1}, \cite{to2} with toroidal magnets that are widely considered as the only convincing experimental verification of the
Aharonov-Bohm effect.

The physical interpretation of the results of the Tonomura et al. experiments is based on the validity of the Ansatz of Aharonov-Bohm  \cite{ab} that is
an approximate solution to the Schr\"odinger equation. Aharonov-Bohm propose a solution to the Schr\"odinger equation when, to a good aproximation,
the electron stays in a simply connected region region of space, $\mathcal C$ (more precisely in a region with trivial first group of singular homology),
where the electromagnetic field is zero. Aharonov-Bohm point out that in
this region the magnetic potential is the gradient of a scalar function, $\lambda(x)$, and that the solution can be found by means of a change of gauge from the free evolution.
The chosen scalar function depends on the simply connected region and it is only defined there. We now state the Aharonov-Bohm Ansatz in a precise way.

\begin{definition}{Aharonov-Bohm Ansatz with  Initial  Condition  at Time Zero}\\
{\rm Let $ A $ be a magnetic potential with curl $A = 0$, defined in a  region $\mathcal C $ that is simply connected, or more precisely with
 trivial first group of singular homology . Let $A =\nabla\lambda(x)$, for some scalar function $\lambda$.  Let $\phi$ be the initial
data at time zero of a solution  to the Schr\"odinger equation that
stays in $ \mathcal C $ for all times, to a good approximation. Then, the change of gauge formula (\cite{ab}, page 487),
\beq
e^{-i t H(A) } \phi \approx  \phi_{AB}(x,t):=
e^{i \lambda(x)}e^{-i tH_0} e^{-i
  \lambda(x)} \phi
\label{4.1}
\ene
holds.}
\label{def-4.2}
\end{definition}
\bull

\noindent  To be more precise, in (\ref{4.1}) we  denote by $\lambda(x)$ an extension of $\lambda(x)$ to a function defined in $\ere^3$.
Note that if the initial state at $t=0$ is taken as
$ e^{-i\lambda(x)} \,\phi$ the Aharonov-Bohm Ansatz is the
multiplication of the free solution by the Dirac magnetic factor
$e^{i \lambda(x) }$ \cite{di}.

Equation (\ref{4.1}) is formulated when the initial conditions are
taken at time zero.  We now find the appropriate  Aharonov-Bohm Ansatz  for the high-velocity solution
\beq
\psi_{\v}= e^{-it H(A,V)}\, W_-(A,V)\, \varphi_{\v},
\label{4.2}
\ene
that satisfies the initial condition at time $-\infty$
\beq
\lim_{t \rightarrow - \infty}\left\|\psi_\v -J \, \psi_{\v,0} \right\|=0,
\label{4.3}
\ene
where $\psi_{\v,0}$ is the free  incoming wave packet that  represents the electron  at the time of emission,
\beq
\psi_{\v,0}:= e^{-itH_0} \, \varphi_\v.
\label{4.4}
\ene
We have to find the initial state at time zero in (\ref{4.1}) in order that the initial condition at time $-\infty$ is satisfied. We take,
$$
\phi = e^{i\lambda(x)}\,  e^{-i\lambda_\infty(-\mo)}\, \varphi_\v,
$$
where, $\lambda_\infty(x):= \lim_{r \rightarrow \infty}\,\lambda(rx)$.   We have that,
$$
 e^{i \lambda(x)}e^{-i tH_0} e^{-i \lambda(x)} \phi= e^{-it H_0} e^{i\lambda(x+(\mo /m) t)}\, e^{-i\lambda_\infty(-\mo)}\, \varphi_\v.
 $$
 But as $\lambda_\infty$ is homogeneous of order zero
 $$
 \hbox{\rm s}-\lim_{t \rightarrow -\infty}\, e^{i\lambda_\infty(x+\mo t)}=  e^{i\lambda_\infty(-\mo)}.
 $$
Then,
$$
\lim_{t\rightarrow -\infty}\, \left\| e^{i \lambda(x)}\,e^{-i tH_0} e^{-i \lambda(x)}\, \phi - e^{-itH_{0}}\, \varphi_\v \right\|=0.
$$
Furthermore, for the high-velocity state $\varphi_\v$ and large $v$ we have that,
\beq
\label{4.5}
 e^{-i \lambda_{\infty}(-\mo)} \varphi_\v \approx
  e^{-i \lambda_{ \infty}(-\hv)}  \varphi_\v.
\ene
For this statement see the proof of Theorem 5.7 of \cite{bw}.
It follows that the Aharonov-Bohm Ansatz for $\psi_\v$ is given by,
$$
\psi_{\v}(x,t) \approx e^{i\lambda(x) } e^{-i t H_0} e^{-i \lambda_{\infty}(-\hv)} \varphi_\v.
$$
We prove below that without loss of generality  we can assume that the potential $A$ has compact support  in $B_R$ and $\lambda_\infty(-\hv)=0$. In
this case the Aharonov-Bohm Ansatz for high-velocity solutions with initial data at time $-\infty$ is given by the following definition.
\begin{definition}{Aharonov-Bohm Ansatz with Initial condition at Time  Minus Infinite}\\
\label{AB-Ansatz}
{\rm Let $ A $ be a magnetic potential with curl $A = 0$, defined in a  region $\mathcal C $ with trivial first group of singular
homology. Let $A =\nabla \lambda(x)$ for some scalar function $\lambda$ with $\lambda_\infty(-\hv)=0$ for some unit vector $\hv$.
Let $ \psi_{\v}(x,t):= e^{-i\frac{t}{\hbar} H(A)} W_{-}(A,V) \,\varphi_{\v} $ be the solution to the Schr\"odinger equation that behaves
like $\psi_{\v,0}:=e^{-i t H_{0}}\varphi_{\v}$ when  time goes to minus infinite.
We suppose that $ \psi_{\v} $ is approximately localized for all times in $ \mathcal C $. Then, the following change of gauge formula holds,
\beq
\label{4.6}
\psi_{\mathbf v} \approx  \psi_{AB, \mathbf v}(x,t):=
e^{i\lambda(x) } e^{-i t H_0} \varphi_\mathbf v.
\ene
}
\end{definition}

\bull

\noindent Observe that, again,  the Aharonov-Bohm Ansatz is the
multiplication of the free solution by the Dirac magnetic factor
$e^{i \lambda(x) }$ \cite{di}.

Note that for the validity of the Aharonov-Bohm Ansatz it is
necessary that the electron stays in the simply connected region $\mathcal C$  (disjoint from the magnet) and that it is not directed towards the magnet
$K$ (it does not hit it). In fact, if the electron hits  $K$ it will be reflected no
matter how big the velocity is, and then, it will not follows the
free evolution multiplied by a phase, as is the case in the
Aharonov-Bohm Ansatz. This can be seen, for example,
in the case of a solenoid contained inside an infinite cylinder, that has explicit solution \cite{ru}. See for example equation (4.22)
of \cite{ru} that gives the phase shifts in the case with Dirichlet boundary condition, that shows that the scattering from the cylinder is always
present and that it appears in the leading order together with the contribution of the magnetic flux inside the cylinder. In fact, the magnet
 $K$  amounts to an infinite electric potential. Observe, however,
that, as we prove below, a finite potential $V$ that satisfies (\ref{2.4}) produces a
lower order term  and, hence, it does not affect the validity of the
Aharonov-Bohm Ansatz for high velocity.

Recall that the  set
$\Lambda_{\hv}$ (\ref{3.24}) corresponds to trajectories that do not
hit the magnet under the classical free evolution. Since for high
velocities the electron follows the quantum free evolution  and as
the quantum free evolution is concentrated along the classical
trajectories, it is natural to require that when the electron is
inside $B_R$  it is actually in $\Lambda_{\hv} \cap B_R$,  in such a
way  that as it crosses the region where the magnet is located it
does so through the holes of $K$ that are in $\Lambda_{\hv}$ or that
it crosses outside of the holes of $K$. In general, $\Lambda_{\hv}$
crosses several holes of $K$ and if two electrons cross different
holes of $K$ there can be no simply connected region that contains
both of them for all times.

In order to make the idea above
precise we have first to decompose $\Lambda_{\hv}$ on its components
that cross the same holes of $K$. This was accomplished in \cite{bw}
as follows.

Suppose that $  L(x,\hv) \subset \Lambda$, and
$L(x,\hv) \cap B_R \neq \emptyset$. we denote by  $c(x,\hv)$ the curve consisting of the segment
$L(x,\hv) \cap \overline{B_R}$ and  an arc on $ \partial\overline{ B_R}$ that connects the points
$L(x,\hv) \cap \partial  \overline{ B_R}$. We orient $c(x,\hv)$ in such a way that the segment of straight
line has the orientation of $\hv$. See Figure 2.

\begin{definition}\label{def-4.3}{\rm
A line  $L(x,\hv) \subset \Lambda$ goes through  holes of $K$ if $L(x,\hv) \cap B_R \neq \emptyset$
and $[c(x,\hv)]_{H_1(\Lambda;\ere)}\neq 0$. Otherwise we say that $L(x,\hv)$ does not go through  holes of $K$.}
\end{definition}
Note that this characterization of lines that go or do not go through  holes of $K$ is independent of the
$R$ that was used in the definition. This follows from the homotopic invariance of homology. See Theorem 11.2,
page 59 of \cite{gh}.

In an intuitive sense $[c(x,\hv)]_{H_1(\Lambda;\ere)}= 0$ means that
$c(x,\hv)$ is the boundary  of a surface (actually of a chain) that is contained in $\Lambda$ and then
it can not go through  holes of $K$. Obviously, as $K \subset  B_R$,  if
$L(x,\hv) \cap B_R = \emptyset$ the line $L(x,\hv)$ can not go through  holes of $K$.

\begin{definition} \label{def-4.4}{\rm
Two lines $L(x,\hv), L(y,\hw) \subset \Lambda$ that go through  holes of $ K $  go through the same holes if $[c(x,\hv)]_{H_1(\Lambda;\ere)}=
\pm [c(y,\hw)]_{H_1(\Lambda;\ere)}$. Furthermore, we say that the lines go through the holes in the same direction
if $[c(x,\hv)]_{H_1(\Lambda;\ere)}=
 [c(y,\hw)]_{H_1(\Lambda;\ere)}$.}
 \end{definition}

\
\begin{remark}\label{remm-4.5}
{\rm
If $(x,\hv)\in \Lambda \times \ese^2$, there are neighborhoods $ B_x \subset \ere^3, B_{\hv}\subset \ese^2$
such that $(x,\hv) \in B_x\times B_{\hv}$ and if $(y,\hw) \in B_x\times B_{\hv}$ then, the following is true:
if $L(x,\hv )$ does not go true  holes of $K$, then, also $L(y,\hw )$ does not go through  holes of $K$.
If $L(x,\hv)$ goes through  holes of $K$, then, $L(y,\hw)$ goes through the same holes and in
the same direction. This follows from the homotopic invariance of homology, Theorem 11.2, page 59 of
\cite{gh}.}
\end{remark}

\begin{definition} \label{def-4.6}{\rm
For any $\hv \in \ese^2$ we denote by $ \Lambda_{\hv, \hbox{\rm out}}$ the set of points
$x \in \Lambda_{\hv}$ such that $L(x,\hv)$ does not go through  holes of $K$. We call this set the region
without holes of $\Lambda_{\hv}$.  The holes of $\Lambda_{\hv}$ is the set $\Lambda_{\hv, \hbox{\rm in}}:=
\Lambda_{\hv} \setminus  \Lambda_{\hv, \hbox{\rm out}} $.}
\end{definition}

\bull

We define the following equivalence relation on $\Lambda_{\hv, \hbox{\rm in}}$. We say that
$ x R_{\hv} y$ if and only if $L(x,\hv)$ and $L(y,\hv)$ go through the same holes and in the same direction. By $[x]$ we designate the
classes of equivalence under $R_{\hv}$. We denote by $\left\{ \Lambda_{\hv, h} \right\}_{h \in \mathcal I}$ the partition of
$\Lambda_{\hv,\hbox{\rm in}}$ given by this equivalence relation. It is defined as follows.
$$
\mathcal I := \{ [x]  \}_{x\in \Lambda_{\hv, \hbox{\rm in}}}.
$$

Given $ h \in \mathcal I$ there is $x \in \Lambda_{\hv, \hbox{\rm in}}$ such that $h=[x]$. We denote,
$$
\Lambda_{\hv,h}:= \{y \in \Lambda_{\hv,\hbox{\rm in}}: y R_{\hv} x  \}.
$$
Then,
$$
\ds
\Lambda_{\hv, \hbox{\rm in}}= \cup_{h \in \mathcal I} \Lambda_{\hv,h},\,\,\,\,  \Lambda_{\hv,h_1}\cap
\Lambda_{\hv,h_2}= \emptyset, \,h_1 \neq h_2.
$$
We call $\Lambda_{\hv,h}$   the subset  of  $\Lambda_{\hv}$  that goes through the holes  $h$ of $K$  in the direction of $\hv$.
Note that
\beq
\{\Lambda_{\hv,h}\}_{h \in \mathcal I} \cup \{\Lambda_{\hv, \hbox{\rm out}}\}
\label{4.7}
\ene
is an  disjoint open cover of $\Lambda_{\hv}$.

We visualize the dynamics of the electrons that travel through the holes of $K$ in $\Lambda_{\hv,h}$ as follows. For large negative times the
incoming electron wave packet is in $\Lambda$,  far away from $K$. As time increases the electron travels towards   $K$  and it reaches  the region where
$K$ is located, let us say that it is inside $B_R$. As these times the electron   has to be in  $\Lambda_{\hv,h}$ in order cross $B_R$ through the holes
of $K$ in $\Lambda_{\hv,h}$. After crossing the holes it travels again away from $K$
towards spatial infinity in $\Lambda$. This means that the classical trajectories have to be in the following domain,

\beq
\mathcal C_h:= \left[\Lambda\setminus \left( \overline{B_R} \cup P_{\hv}  \right)\right] \cup \left( \overline{B_R} \cap\Lambda_{\hv,h}\right),
\label{4.8}
\ene

where $P_{\hv}$ is the plane orthogonal to $\hv$ that passes through zero,
\beq
P_{\hv}:=\left\{ x \in \ere^3: x\cdot \hv=0      \right\}.
\label{4.9}
\ene
Note that we take away from $\mathcal C_h$   the part of $P_{\hv}$ that does not intersects $\Lambda_{\v,h}$ in order that the only way that the electron
in $\mathcal C_h$ can classically cross the plane  $P_{\hv}$ is through $\Lambda_{\hv,h}$.

In a similar way, the classical trajectories of the electrons that  do not cross any hole of $K$ have to be on the set
 \beq
\mathcal C_{\rm out}:=\left( \Lambda \setminus \overline{B_R}\right) \cup \left( \overline{B_R} \cap\Lambda_{\hv, \rm out}  \right).
\label{4.10}
\ene

In  Corollary \ref{cor-a.9} in the  appendix we prove that that the first group of singular homology with coefficients in $\ere$ of $\mathcal C_h,
H_1(\mathcal C_h; \ere), h \in \mathcal I $, and of $\mathcal C_{\rm out},H_1(\mathcal C_{\rm out}; \ere)$ are
trivial. We actually prove that the first de Rham cohomology class of  $\mathcal C_h$ and of $\mathcal C_{\rm out}$ are trivial by explicitly constructing
a function $\lambda$ such that $A= \nabla\lambda$ for any magnetic potential $A$ with ${\rm curl}\, A=0$, or
 in differential geometric language by
constructively proving that any closed one form is exact. Then, the triviality of the  the first group of singular homology with coefficients in $\ere$
of $\mathcal C_h$ and of $\mathcal C_{\rm out}$  follows from de Rham's theorem (Theorem 4.17 page 154 of \cite{w}).

Let $x_0$ be a fixed point with $x_0 \cdot \hv < - R$. We define,
\beq
\lambda_{h }(x):= \int_{C^h}\, A , \quad h \in \mathcal I, \,\hbox{\rm where $C^h$ is any differentiable path from $x_0$ to $x$ in $\mathcal C_h$},
\label{4.11}
\ene
and,

\beq
\lambda_{\rm out}(x):= \int_{C^{{\rm out}}}\, A , \quad \hbox{\rm where $C^{\rm out}$ is any differentiable path from $x_0$ to $x$ in
$\mathcal C_{\rm out}$}.
\label{4.12}
\ene
Since $H_1(\mathcal C_h; \ere), h \in \mathcal I $ and $H_1(\mathcal C_{\rm out}; \ere)$ are trivial, $\lambda_h, h \in \mathcal I$ and
$\lambda_{\rm out}$ do not depend  in the particular curve form $x_0$ to $x$ that we take, respectively, in $ \mathcal C_h, h \in \mathcal I$
 and $ \mathcal C_{\rm out}$. Furthermore, they are differentiable and $\nabla \lambda_h(x)= A(x), x \in \mathcal C_h , h \in \mathcal I$ and
$ \nabla \lambda_{\rm out}(x)= A(x), x \in \mathcal C_{\rm out}$.

Before we prove the validity of the Aharonov-Bohm Ansatz we  prepare some simple results on the free evolution that we need.
Below we denote by $\widetilde{O}$ the complement of any set $O \subset \ere^3$.
\begin{lemma} \label{lemm-4.7}
We denote,
\beq\label{4.13}
C_{-,h}:= \{ x \in \Lambda \setminus \overline{B_R} : x\cdot \hv < 0  \} \cup \Lambda_{\hv,h}, h \in \mathcal I,
\quad C_{-, \rm out}:=  \{ x \in \Lambda \setminus \overline{B_R} : x\cdot \hv < 0  \}\cup \Lambda_{\hv,\rm out}.
\ene
Then,  for any $l=0,1, \cdots$ and any compact set $\Lambda_0 \subset \Lambda_{\hv,h}, h \in \mathcal I$       there is a constant $C_l$ such that
 $ \forall Z \geq 0, \forall z \in (-\infty, Z]$, and for all $\varphi \in \mathcal H_2(\ere^3)$ with support in $\Lambda_0$,
\beq
\label{4.14}
\left\| \chi_{\ds \widetilde{C_{-,h}}} \, e^{-i\frac{z}{v}H_0}\, \varphi_{\v} \right\|_{L^2(\ere^3)}\leq C_l\, \left( (1+Z)^{-l}+ \frac{1+Z}{v}\right)
\|\varphi\|_{\mathcal H_2(\ere^3)}.
\ene
Furthermore,  for any $l=0,1, \cdots$ and any compact set $\Lambda_0 \subset \Lambda_{\hv,\rm out}$ there is a constant $C_l$ such that
 $ \forall Z \geq 0, \forall z \in (-\infty, Z]$, and for all $\varphi \in  \mathcal H_2(\ere^3)$ with support in $\Lambda_0$,
\beq
\label{4.15}
\left\| \chi_{\ds \widetilde{C_{-,\rm out}}} \, e^{-i\frac{z}{v}H_0}\, \varphi_{\v} \right\|_{L^2(\ere^3)}\leq C_l\, \left( (1+Z)^{-l}+ \frac{1+Z}{v}\right)
\|\varphi\|_{\mathcal H_2(\ere^3)}.
\ene
\end{lemma}

\noindent{\it Proof:} We give the proof of (\ref{4.14}). Equation (\ref{4.15}) follows in the same way.
\begin{enumerate}
\item
Suppose that $ z \leq \, {\rm min} \,( -\frac{4}{3} R, -Z)$. By (\ref{3.16}) it is enough to prove (\ref{4.14}) for $\tilde{\varphi}$. The
estimate follows from (\ref{3.9}) and Lemma \ref{lemm-3.2} observing that $\chi_{\ds \widetilde{C_{-,h}}}(x)= \chi_{\ds \widetilde{C_{-,h}}}(x)
F( |x-z\hv| > |z|/4 )$.
\item
Suppose that $ z \in [-Z,Z]$. Since, $ \chi_{\ds \widetilde{C_{-,h}}} \, e^{\ds-i z \mo\cdot\hv}\, \varphi =0$, it follows from (\ref{3.8})
that,
\beq
\label{4.16}
\begin{array}{l}
\left\| \chi_{\ds\widetilde{C_{-,h}}} \, e^{-i\frac{z}{v}H_0}\, \varphi_{\v} \right\|_{L^2(\ere^3)}=
\left\| \chi_{\ds\widetilde{C_{-,h}}} \,\left[ e^{-i z H_1}- e^{\ds-i z \mo\cdot\hv}e^{\ds -i z m v/2}\right] \, \varphi
 \right\|_{L^2(\ere^3)} \leq \\\\
 C \frac{Z}{v}\, \|\varphi\|_{\mathcal H_2(\ere^3)}.
\end{array}
 \ene
 \item If $ Z \leq \frac{4}{3} R$ it remains to consider $ z \in [-\frac{4}{3} R, -Z]$. In this case we just say that,
\beq
\label{4.17}
\left\| \chi_{\ds\widetilde{C_{-,h}}} \, e^{-i\frac{z}{v}H_0}\, \varphi_{\v} \right\|_{L^2(\ere^3)} \leq \|\varphi\|_{L^2(\ere^3)}
 \leq C_l (1+Z)^{-l }\, \|\varphi\|_{L^2(\ere^3)}.
 \ene
\end{enumerate}
\begin{lemma} \label{lemm-4.8}
We denote,
\beq\label{4.18}
 C_+^0:= \{ x \in \Lambda \setminus \overline{B_R} : x\cdot \hv > 0  \}.
\ene
Then,  for any $l=0,1, \cdots$ there is a constant $C_l$ such that
 $ \forall Z \geq 0, \forall z \geq  Z$, and for all $\varphi \in \mathcal H_2(\ere^3)$,
\beq
\label{4.19}
\left\| \chi_{\ds\widetilde{C_{+}^0}} \, e^{-i\frac{z}{v}H_0}\, \varphi_{\v} \right\|_{L^2(\ere^3)}\leq C_l\, \left( (1+Z)^{-l}+ \frac{1}{v}\right)
\|\varphi\|_{\mathcal H_2(\ere^3)}.
\ene
\end{lemma}

\noindent{\it Proof:}
If  $ Z \geq \, \frac{4}{3} R$ we prove (\ref{4.19}) as in item 1 of the proof of Lemma \ref{lemm-4.7} observing that $\chi_{\ds \widetilde{C_{+}}}(x)=
 \chi_{\ds \widetilde{C_{+}}}(x)
F( |x-z\hv| > |z|/4 )$.  If $ Z \leq \frac{4}{3} R$ it remains to consider $ z \in [Z, \frac{4}{3} R ]$ but in this case (\ref{4.19}) follows as in item 3
 of the proof of Lemma \ref{lemm-4.7}.

\begin{corollary} \label{cor-4.9}
For any $l=0,1, \cdots$ and any compact set $\Lambda_0 \subset \Lambda_{\hv,h}, h \in \mathcal I$       there is a constant $C_l$ such that
 $ \forall Z \geq 0, \forall z \in \ere$, and for all $\varphi \in \mathcal H_2(\ere^3)$ with support in $\Lambda_0$,
\beq
\label{4.20}
\left\| \chi_{\ds\widetilde{\mathcal C_h}} \, e^{-i\frac{z}{v}H_0}\, \varphi_{\v} \right\|_{L^2(\ere^3)}\leq C_l\, \left( (1+Z)^{-l}+ \frac{1+Z}{v}\right)
\|\varphi\|_{\mathcal H_2(\ere^3)}.
\ene
Furthermore,  for any $l=0,1, \cdots$ and any compact set $\Lambda_0 \subset \Lambda_{\hv,\rm out}$ there is a constant $C_l$ such that
 $ \forall Z \geq 0, \forall z \in \ere$, and for all $\varphi \in  \mathcal H_2(\ere^3)$ with support in $\Lambda_0$,
\beq
\label{4.21}
\left\| \chi_{\ds \widetilde{\mathcal C_{\rm out}}} \, e^{-i\frac{z}{v}H_0}\, \varphi_{\v} \right\|_{L^2(\ere^3)}\leq C_l\,
\left( (1+Z)^{-l}+ \frac{1+Z}{v}\right)
\|\varphi\|_{\mathcal H_2(\ere^3)}.
\ene
\end{corollary}
\noindent{Proof :} Note that since
$$
\left(\Lambda\setminus  \overline{B_R}\right) \cap \Lambda_{\hv,h} \subset \left[\Lambda\setminus \left( \overline{B_R} \cup P_{\hv}  \right)\right],
\quad h \in \mathcal I,
$$
we have that,
$$
C_{-,h} \subset \mathcal C_h, h \in \mathcal I.
$$
Moreover,
$$
C_{-,\rm out } \subset \mathcal C_{\rm out},
$$
and,
$$
C_+^0 \subset \mathcal C_h \, \cap \, \mathcal C_{\rm out}.
$$
Hence, the corollary follows from Lemma \ref{lemm-4.7} when $ z \leq Z$ and from Lemma \ref{lemm-4.8} when $z \geq Z$.

\begin{definition}\label{def-4.10}
{\rm
We designate by $\0p2$ the set of all potentials $A \in \p2$ that satisfy,
$$
{\rm curl}\, A =B=0.
$$
}
\end{definition}
\begin{remark}
\label{rem-4.11}
{\rm
For any  $ A \in \0p2 \cap C^l(\overline{\Lambda}, \ere^3), l=1,2,\cdots$ there is a $ \tilde{A}\in \0p2 \cap C^l(\overline{\Lambda}, \ere^3)$
with the same flux as $A$ and with
support$ \tilde{A} \subset B_R$. To prove this statement we take any $x_0 \in \Lambda\setminus B_R$ and let $ \varepsilon >0$ be so small that
$K \subset B_{R-\varepsilon}$. We define,
$$
\overline{\lambda}(x):= \int_{C(x_o,x)}A, \quad {\rm for}\, x \in \Lambda \setminus B_{R-\varepsilon},
$$
where $ C(x_0,x)$ is any differentiable path from $x_0$ to $x$ contained in $\Lambda \setminus B_{R-\varepsilon}$.
Then, $\overline{\lambda} \in C^l(\overline{\Lambda} \setminus B_{R-\varepsilon})$. We denote by $ \lambda$ any extension of $\overline{\lambda}$
to $\ere^3$ such that $\lambda \in C^l(\ere^3)$ \cite{tr}. We define,
$$
\tilde{A}(x):= A(x) - \nabla \lambda(x), \quad x \in \overline{\Lambda}.
$$
Then, $\tilde{A}\in  \0p2 \cap C^l(\overline{\Lambda}, \ere^3), l=1,2,\cdots$, the flux of $\tilde{A}$ is the same as the one of $A$ and  support$ \tilde{A} \subset B_R$.
Note that if $B=0$ the Coulomb potential $A_C \in C^\infty(\overline{\Lambda}, \ere^3)$ (see Theorem 3.7 of \cite{bw}).
Doing the gauge transformation above we see that for every $l=1,2, \cdots$ there is a potential in $\0p2 \cap C^l(\overline{\Lambda}, \ere^3)$
with compact support in $B_R$.
}
\end{remark}
By Remark \ref{rem-4.11} we can use the freedom of taking a gauge transformation to assume that $ A \in \0p2 \cap C^1(\overline{\Lambda}, \ere^3)$
and that support$A \subset B_R$, what we do from now on.

\begin{theorem}\label{theor-4.12}
For any $l=0,1, \cdots$ and any compact set $\Lambda_0 \subset \Lambda_{\hv,h}, h \in \mathcal I$ there is a constant $C_l$ such that
 $ \forall Z \geq 0, \forall z \in \ere$, and for all $\varphi \in \mathcal H_2(\ere^3)$ with support in $\Lambda_0$,
\beq
\label{4.22}
\left\| e^{\ds -i \frac{z}{v} H(A,V)}\, W_-(A,V)\, \varphi_\v - e^{i \lambda_h}\,\chi_{\ds\mathcal C_h} \, e^{-i\frac{z}{v}H_0}\, \varphi_{\v} \right\|_{L^2(\ere^3)}
\leq C_l\, \left( (1+Z)^{-l}+ \frac{1+Z}{v}\right)
\|\varphi\|_{\mathcal H_2(\ere^3)}.
\ene
Furthermore,  for any $l=0,1, \cdots$ and any compact set $\Lambda_0 \subset \Lambda_{\hv,\rm out}$ there is a constant $C_l$ such that
 $ \forall Z \geq 0, \forall z \in \ere$, and for all $\varphi \in  \mathcal H_2(\ere^3)$ with support in $\Lambda_0$,
\beq
\label{4.23}
\left\| e^{\ds -i \frac{z}{v} H(A,V)}\, W_-(A,V)\, \varphi_\v - e^{i\lambda_{\rm out}}\,\chi_{\ds \mathcal C_{\rm out}} \, e^{-i\frac{z}{v}H_0}\, \varphi_{\v} \right
\|_{L^2(\ere^3)}\leq C_l\,
\left( (1+Z)^{-l}+ \frac{1+Z}{v}\right)
\|\varphi\|_{\mathcal H_2(\ere^3)}.
\ene
\end{theorem}

\noindent{\it Proof:} We first consider the case  $z \leq Z$. In this case
the theorem follows from Lemmata \ref{lemm-4.7}, \ref{lemm-4.8}, Corollary \ref{cor-4.9},  and  (\ref{3.59}) observing that  that since
support $A \subset B_R$ ,
$$
 - L_{A, \hv}(-\infty) =  \lambda_h(x), \quad x \in C_{-,h}, h \in \mathcal I, \quad - L_{A, \hv}(-\infty)=   \lambda_{\rm out}(x),
\quad x \in
C_{-,\rm out}.
$$
For $ z \geq Z$ we use (\ref{3.61}), Lemma \ref{lemm-4.8} and Corollary \ref{cor-4.9}. For this purpose note that,
$$
\int_{-\infty}^\infty \, A(x+\tau \hv)\cdot \hv\, d\tau = \int_{c(x,\hv)}\, A, \quad \,{\rm for}\, x \in \Lambda_{\hv,h}, h\in \mathcal I,
\int_{-\infty}^\infty \, A(x+\tau \hv)\cdot \hv\, d\tau=0,\,{\rm for}\,    x \in \Lambda_{\rm out}.
$$
Moreover, recall that (see Definition 7.10 of \cite{bw})
$$
F_h:= \int_{c(x,\hv)}\, A, \quad  x \in \Lambda_{\hv,h}, h \in \mathcal I,
$$
and that $F_h$ is constant for all $ x \in \Lambda_{\hv,h}$. $F_h$  is the magnetic flux over any surface (or a chain) in $\ere^3$
whose boundary is $c(x,\hv)$. In other words, it is the flux associated to the holes of $K$ in $\Lambda_{\hv,h}$ . Furthermore, we have that,
\beq\label{4.23b}
F_h= \lambda_h(x), \quad x \in C_{+}^0,
\ene
what completes the proof for $ z \geq Z, h \in \mathcal I$. For the case $\Lambda_{\hv,\rm out}$ and $ z \geq Z$ we observe that,
\beq\label{4.23c}
\lambda_{\rm out}(x)=0, \quad {\rm for} \, x \in C_+^0.
\ene
We now state our main results on the validity of the Aharonov-Bohm Ansatz.
\begin{theorem}\label{theor-4.13}
For any $ 1 > \delta >0$ and any compact set $\Lambda_0 \subset \Lambda_{\hv,h}, h \in \mathcal I$ there is a constant $C_\delta$ such that
 $ \forall t \in \ere$ and for all $\varphi \in \mathcal H_2(\ere^3)$ with support in $\Lambda_0$,
\beq
\label{4.24}
\left\| e^{\ds -i t H(A,V)}\, W_-(A,V)\, \varphi_\v - e^{i\lambda_h}\,\chi_{\ds\mathcal C_h} \, e^{-itH_0}\, \varphi_{\v} \right\|_{L^2(\ere^3)}
\leq \frac{C_\delta}{v^{1-\delta}}
\|\varphi\|_{\mathcal H_2(\ere^3)}.
\ene
Furthermore,  for any $ 1 > \delta >0$ and any compact set $\Lambda_0 \subset \Lambda_{\hv,\rm out}$ there is a constant $C_\delta$ such that
 $ \forall t \in \ere$ and for all $\varphi \in  \mathcal H_2(\ere^3)$ with support in $\Lambda_0$,
\beq
\label{4.25}
\left\| e^{\ds -i t H(A,V)}\, W_-(A,V)\, \varphi_\v - e^{i \lambda_{\rm out}}\chi_{\ds \mathcal C_{\rm out}} \, e^{-i tH_0}\, \varphi_{\v} \right
\|_{L^2(\ere^3)}\leq \frac{C_\delta}{v^{1-\delta}}
\|\varphi\|_{\mathcal H_2(\ere^3)}.
\ene
\end{theorem}

\noindent{\it Proof:} we take in Theorem \ref{theor-4.12}, $ Z= v^{1/(1+l)}$ and $ t= z/v$ . Then, for $ v >1$,  $ \frac{1}{v} (1+Z)\leq 2 \frac{1}{ v^{1-1/(1+l)}}$
and $(1+Z)^{-l} \leq \frac{1}{v^{1-1/(1+l)}}$. The theorem follows taking $ \frac{1}{1+l}\leq \delta$.

\bull

Let us take any $ \varphi_0 \in  \mathcal H_2(\ere^3)$ with compact support in $\Lambda_{\hv}$. Then, since (\ref{4.7}) is a disjoint open cover of $\Lambda_{\hv}$
\beq
\varphi_0 = \sum_{h \in \mathcal I}\,\varphi_h +\varphi_{\rm out},
\label{4.26}
\ene
where  $\varphi_h, \varphi_{\rm out} \in \mathcal H_2(\ere^3), \varphi_h$ has compact  support in $\Lambda_{\hv,h}, h \in \mathcal I,$ and
$ \varphi_{\rm out}$ has compact support in $ \Lambda_{\hv, \rm out}$. The sum is finite because $\varphi_0$ has compact support.
We denote,
\beq
\varphi_{\v}:= e^{im \v\cdot x}\, \varphi_0, \, \varphi_{\v,h}:= e^{im\v\cdot x}\, \varphi_h, h \in \mathcal I, \varphi_{\v,\rm out}:=
e^{im \v\cdot x}\, \varphi_{\rm out}.
\label{4.27}
\ene
We define,
\beq
\psi_{AB,\v,h}:= \chi_{\mathcal C_h} \, e^{i\lambda_h}\, e^{-it H_0}\,\varphi_{\v,h}, h \in \mathcal I, \quad
\psi_{AB,\v,\rm out}:= \chi_{\mathcal C_{\rm out}} \, e^{i\lambda_{\rm out}}\, e^{-it H_0}\,\varphi_{\v,\rm out},
\label{4.28}
\ene
\beq
\psi_{AB,\v}:= \sum_{h \in \mathcal I}\, \psi_{AB,\v,h}+ \psi_{AB,\v,\rm out}.
\label{4.29}
\ene
Equation (\ref{4.29}) gives the Aharonov-Bohm Ansatz in the domain $ \cup_{h \in \mathcal I}  \mathcal C_h  \cup \mathcal C_{\rm out}$
 that has non-trivial first group of singular homology as the sum of the Aharonov-Bohm Ans\"atze in each of the components,
 $\mathcal C_h, h \in \mathcal I, \mathcal C_{\rm out}$ that have trivial first group of singular homology.
As we already mentioned, for the Ansatz of Aharonov-Bohm to be valid, it is necessary that the electron
does not hit the magnet. Otherwise, the electron will be reflected and the Ansatz cannot be an
approximate solution because  it  consists of the free evolution  multiplied by a phase  in  configuration space.
Hence, the wave function that represents such an electron has to have its support approximately contained
for  all  times in the domain $ \cup_{h \in \mathcal I}  \mathcal C_h  \cup \mathcal C_{\rm out}$.  In the next theorem we prove
that the Ansatz of Aharonov-Bohm is actually valid on  the biggest domain  where it can be valid, $ \cup_{h \in \mathcal I}
\mathcal C_h  \cup \mathcal C_{\rm out}$, and, in this way, we provide an approximate solution for all times for every electron that does not hit
the magnet.

\begin{theorem}\label{theor-4.14} The Validity of the Aharonov-Bohm Ansatz.

For any $ 1 > \delta >0$ and any compact set $\Lambda_0 \subset \Lambda_{\hv}$ there is a constant $C_\delta$ such that
 $\forall t \in \ere$ and for all $\varphi \in \mathcal H_2(\ere^3)$ with support in $\Lambda_0$ the solution to the Schr\"odinger equation
$e^{\ds -it H(A,V)}\, W_-(A,V)\, \varphi_\v$ that behaves as $ e^{-it H_0}\, \varphi_{\v}$ as $t \rightarrow -\infty$ is given at time $t$ by the
Aharonov-Bohm Ansatz, $\psi_{AB,\v}$, up to the following error,
\beq
\label{4.30}
\left\| e^{\ds -it H(A,V)}\, W_-(A,V)\, \varphi_\v - \psi_{AB,\v} \right\|_{L^2(\ere^3)}
\leq \frac{C_\delta}{v^{1-\delta}}
\|\varphi\|_{\mathcal H_2(\ere^3)}.
\ene
\end{theorem}
\noindent{\it Proof:} The theorem follows from Theorem \ref{theor-4.13}  and equations  (\ref{4.26} to  \ref{4.29}).

\bull

Note that by (\ref{4.23b}, \ref{4.23c}) behind the magnet in $C^0_+$,

\beq\label{4.31}
\psi_{AB,\v,h}:= \chi_{\mathcal C_h} \, e^{iF_h}\, e^{-it H_0}\,\varphi_{\v,h}, h \in \mathcal I, \quad x \in C^0_+,
\ene
and that,
\beq\label{4.32}
\psi_{AB,\v,\rm out}:= \chi_{\mathcal C_{\rm out}} \, e^{-it H_0}\,\varphi_{\v,\rm out}, \quad  \quad x \in C^0_+.
\ene
As mentioned in the introduction the phase shifts $e^{iF_h}$ were measured in the experiments of
Tonomura et al. \cite{to3,to1,to2}
and, furthermore,  since  the Aharonov-Bohm Ansatz is free evolution, up to a phase, the electron is not accelerated, what explains the results of the
experiment of Caprez et al. \cite{cap}. Hence,  Theorem \ref{theor-4.14} rigorously proves that quantum mechanics predicts the results of the
experiments of Tonomura et al. and of Caprez et al..

\section{Appendix}\sss
In this appendix we prove that the first  group of singular homology with coefficients in $\ere$ of $\mathcal C_h$ and of $\mathcal C_{\rm out}$ are
trivial. The sets $\mathcal C_h$ and $\mathcal C_{\rm out}$ are defined, respectively, in (\ref{4.8}) and (\ref{4.10}). We denote
\beq
C_{+}:= \{ x \in \ere^3 \setminus B_R : x\cdot \hv > 0  \}, \quad C_{-}:= \{ x \in \ere^3 \setminus B_R : x\cdot \hv < 0  \},
\label{5.1}
\ene
and by $C^0_{\pm}$ the interior of  $C_{\pm}$. Recall that $P_{\hv}$ is defined in (\ref{4.9}). Then,

\beq
\mathcal C_h= C^0_- \cup C^0_+ \cup (\overline{B_R} \cap \Lambda_{\hv,h}),
\label{5.2}
\ene

\beq
\mathcal C_{\rm out}= C^0_- \cup C^0_+ \cup (\overline{B}_R \cap \Lambda_{\hv,\rm out})\cup (P_{\hv} \setminus \overline{B}_R).
\label{5.3}
\ene
We first prepare several results that we need.
Below we denote by $A$ any continuously differentiable  vector field defined, respectively, in $\mathcal C_h, h \in \mathcal I$, and in $ \mathcal C_{\rm out}$,
with $ {\rm curl}\, A=0$.

Let $x_0$ be a fixed point with $x_0  < - R$. For any $x \in B_R$ we denote, respectively by $\xin, \xout$ the intersection of the line
$\{ x+\tau \hv, \tau \in \ere \}$ with $\partial B_R$ such that $ \xin\cdot \hv < 0, \xout \cdot\hv >0$. For any $ h \in \mathcal I$ let $x^h$ be a
fixed point
in $ \Lambda_{\hv,h} \cap B_R$ and let $\+out$ be a fixed point in $ \Lambda_{\hv, \rm out} \cap B_R $.

\begin{remark}
\label{remm-a.2}{\rm
For every $ x \in C_-$ we denote by $C_-^x$ any differentiable path in $C_-$ that goes from $x_0$ to $x$ and we define,
\beq
\lambda_-(x):=\int_{C_-^x}\, A.
\label{5.4}
\ene
Since $C_-$ is simply connected the line integral in (\ref{5.4}) does not depend on the particular curve $C_-^x$ that we choose.
Then,  for $x \in C_-^0, \,\lambda_-(x)$ is differentiable and $ \nabla \lambda_-(x)= A(x)$.}
\end{remark}

\begin{remark}\label{remm-a.3}{\rm
For every $x \in B_R \cap \Lambda_{\hv}$  we denote by  $C^x_0$  the differentiable  path consisting of a path $C_-^{\xin}$ followed by the segment
$[\xin ,x]$ and we define for every $ x \in   B_R \cap \Lambda_{\hv}$,
\beq\label{5.5}
\lambda_0(x):= \int_{C^x_0}\, A.
\ene
By Remark \ref{remm-a.2} the line integral in (\ref{5.5}) does not depend on the particular curve $C_-^{\xin}$ that we choose.
Then,  for $x \in  B_R \cap \Lambda_{\hv}$ , $\lambda_0(x)$ is differentiable and $ \nabla \lambda_0(x)= A(x)$. To prove this statement we observe that
for each $x \in  B_R \cap \Lambda_{\hv} $ there is $ \varepsilon >0$ such that $B_{\varepsilon}(x) \subset  B_R \cap \Lambda_{\hv}$. The set
$C_s:=\{ C_- \cup (B_{\varepsilon}(x)+ \ere \hv) \}$ is simply connected and, furthermore, $\lambda_0(x)= \int_C\, A$ where $C$ is any differentiable
path contained in $C_s$ that goes from $x_0$ to $x$.
}
\end{remark}
\begin{remark}
\label{remm-a.4}
{\rm
For every $x\in C_+$ and any $ h \in \mathcal I$ we denote by $C_{h,+}^x$ a differentiable path consisting of any curve
$\ds C_-^{x^h_{\rm in}}$ followed from the segment $[x^h_{\rm in}, x^h_{\rm out}]$ and of a differentiable path $C_+^{x^h_{\rm out},x}$ in $C_+$.
The differentiable path $ C_{\rm out,+}^x$ is defined in the same way, but replacing $x^h$ by $\+out$.  We define,
\beq
\lambda_+^h(x):= \int_{C_{h,+}^x}\, A, \quad x \in C_+, h \in \mathcal I,
\label{5.6}
\ene
and

\beq
\lambda_+^{\rm out}(x):= \int_{C_{\rm out,+}^x}\, A, \quad x \in C_+.
\label{5.7}
\ene
Since $C_\pm$ are simple connected $\lambda_+^h$ does not depends of the particular paths $\ds C_-^{x^h_{\rm in}}, C_+^{x^h_{\rm out},x}$ that we choose
and, $\lambda_+^{\rm out}$  does not depends of the particular paths $\ds C_-^{x^{\rm out}_{\rm in}}, C_+^{x^{\rm out}_{\rm out},x}$ that we choose .
It follows that $ \lambda_+^h  $ and $\lambda_+^{\rm out}$ are continuously differentiable in $C_+^0$ and that $\nabla \lambda_+^h(x)=A(x)$,
$ \nabla\lambda_+^{\rm out}(x)= A(x)$.}
\end{remark}
\begin{remark}
\label{remm-a.5}
{\rm
$ \lambda_+^h, h \in \mathcal I$ does not depend of the particular $ x^h \in \Lambda_{\hv,h}$ that we choose. To prove this statement let us take any
$y \in  \Lambda_{\hv,h} \cap B_R$ and let the differentiable path $ C_{y,+}^x$ be defined as  $C_{h,+}^x$ but with $y$ instead of $x^h$. Let $\gamma$
be any differentiable path from  $x$ to $x_0$ contained in $\Lambda \setminus B_R$. Let $C$ be a the closed oriented  differentiable path
consisting of $C_{h,+}^x$, from $x_0$ to $x$, followed from $\gamma$. $C_y$ is defined in the same way, but with $C_{y,+}^x$ instead of $C_{h,+}^x$.
Let $D$ be an arc on $\partial B_R$ from $ x_{\rm in}^h$ to $x_{\rm out}^h$ and let $G$ be a differentiable path  consisting of $\ds C_-^{x^h_{\rm in}}$
followed of $D, C_+^{x^h_{\rm out},x}$ and  $\gamma$. Since $\ere^3 \setminus B_R$ is simply connected we have that,
$$
\int_G\, A =0,
$$
and then,
$$
\int_C\, A= \int_{c(x^h_{\rm in}, \hv)}\, A.
$$
We prove in the same way that,
$$
\int_{C_y}\, A= \int_{c(y_{\rm in}, \hv)}\, A.
$$
Furthermore, since $x^h,y \in \Lambda_{\hv,h}$, we have that $[c(x^h_{\rm in},\hv)]_{H_1(\Lambda;\ere)}=
 [c(y_{\rm in},\hv)]_{H_1(\Lambda;\ere)}$, and then, by Stoke's theorem,
$$
 \int_{c(x^h_{\rm in}, \hv)}\, A = \int_{c(y_{\rm in}, \hv)}\, A,
$$
what proves that,
$$
\int_C\, A= \int_{C_y}\, A,
$$
and then,
$$
\lambda_+^h(x):= \int_{C_{h,+}^x}\, A = \int_{C_{y,+}^x}\, A.
$$
}
\end{remark}
\begin{remark}\label{remm-a.6}
{\rm $ \lambda_+^{\rm out}$ does not depend of the particular $ x^{\rm out} \in \Lambda_{\hv,\rm out}$ that we choose. This is proven as in
Remark \ref{remm-a.5} replacing $x^h$ by $x^{\rm out}$. Furthermore, as in this case $[c(x^{\rm out}
_{\rm in},\hv)]_{H_1(\Lambda;\ere)}= 0$,
$$
\int_C\, A= 0,
$$
and then,
\beq
\lambda_+^{\rm out}(x)= \int_{\gamma}\, A,
\label{5.8}
\ene
where $ \gamma$ is any  differentiable path from $x_0$ to $x$ contained in $\Lambda \setminus B_R$.
}
\end{remark}
\begin{definition}
\label{def-5.7}
{\rm
For all $ h \in \mathcal I$ we  define  $ \lambda^h: \mathcal C_h \rightarrow \ere$ as follows,
\beq
\label{5.9}
\lambda^h(x):= \left\{
\begin{array}{l}
\lambda_-(x), \quad \hbox{\rm if}\,\, x \in C_-, \\\\
\lambda_0(x), \quad \hbox{\rm if}\,\, x \in  \Lambda_{\hv,h}\cap B_R,\\\\
\lambda_+^h(x), \quad \hbox{\rm if}\,\, x \in C_+.
\end{array}
\right.
\ene
Furthermore, we define $\lambda^{\rm out}: \mathcal C_{\rm out} \rightarrow \ere$ as,
\beq
\label{5.10}
\lambda^{\rm out}(x):= \left\{
\begin{array}{l}
\lambda_-(x), \quad \hbox{\rm if}\,\, x \in C_- ,\\\\
\lambda_0(x), \quad \hbox{\rm if}\,\, x \in  \Lambda_{\hv,\rm out}\cap B_R,\\\\
\lambda_+^{\rm out}(x), \quad \hbox{\rm if}\, x \in C_+,\\\\
\int_{\gamma}\, A,  \hbox{\rm if \, $x \in P_{\hv}\setminus B_R$, where $\gamma$ is any differentiable path from $x_0$ to $x$
contained in $\Lambda \setminus B_R$
 }.
\end{array}
\right.
\ene
}
\end{definition}
\begin{lemma}
\label{lemm-a.8}
The functions $\lambda^h, h \in \mathcal I$ and $\lambda^{\rm out}$ are continuously differentiable and
$ \nabla \lambda^h(x)= A(x), x \in \mathcal C_h, h \in \mathcal I$ and $\nabla \lambda^{\rm out}(x)= A(x), x \in \mathcal C^{\rm out}$.
\end{lemma}

\noindent{\it Proof:} We first consider  $\lambda^h, h \in \mathcal I$. By Remarks \ref{remm-a.2}, \ref{remm-a.3} and \ref{remm-a.4} $\lambda^h(x)$
 is continuously
differentiable and $ \nabla \lambda^h(x)= A(x)$ for $x \in C_-^0 \cup C_+^0 \cup \Lambda_{\hv,h}\cap B_R$. If follows from (\ref{5.2}) that it only
remains to prove
the result for $ x \in \Lambda_{\hv,h} \cap \partial B_R$. Let $\varepsilon > 0$ be such that, $B_{\varepsilon}(x) \subset \Lambda_{\hv,h}$
(see Remark \ref{remm-4.5}). The set
$$
C_{p,h}:= \{C_-^0 \cup (B_{\varepsilon}(x)+ \ere \hv) \cup C_+^0 \}
$$
is simply connected and by Remark \ref{remm-a.5}
$$
\lambda^h(y) = \int_C\, A, \quad y\in C_{p,h},
$$
where $C$ is any differentiable path from $x_0$  to y that is contained in $C_{p,h}$. It follows that $\lambda^h(x)$ is differentiable for
 $x \in \Lambda_{\hv,h} \cap \partial B_R$ and that $\nabla \lambda^h(x)= A(x)$.

 Let us now consider $\lambda^{\rm out}$. By Remarks \ref{remm-a.2}, \ref{remm-a.3} and \ref{remm-a.4} the lemma holds for
 $ x \in  C_-^0 \cup C_+^0 \cup \left(\Lambda_{\hv,\rm out}\cap B_R\right)$. Furthermore, by the definition of $\lambda^{\rm out}$ and (\ref{5.8}) it also holds for
 $x \in P_{\hv}\setminus \overline{B_R}$. By (\ref{5.3}) it only remains to consider the case of $ x \in \partial B_R \cap \Lambda_{\hv,\rm out}$.
Take $\varepsilon >0$ such that $ K \subset B_{R-\varepsilon}$. Then, since $\ere^3\setminus \overline{B_{R-\varepsilon}}$ is a simply connected set where
${\rm curl} A=0$  we have that  for $ x \in \mathcal C_{\rm out} \setminus \overline{B_{R-\varepsilon}}$
$$
\lambda^{\rm out}(x)=  \int_{\gamma}\, A,
$$
where $\gamma$ is any differentiable path from $x_0$ to $x$
contained in $\ere^3 \setminus \overline{B_{R-\varepsilon}}$.
This implies that $\lambda^{\rm out}(x)$ is continuously differentiable with $\nabla \lambda^{\rm out}(x)= A(x)$ for
$x \in \mathcal C^{\rm out} \setminus \overline{B_{R-\varepsilon}}$ and in particular for $ x \in \partial B_R \cap \Lambda_{\hv.\rm out}$.

\begin{lemma}\label{lemm-a.1}
The first de Rham cohomogoly groups $H^1_{de R}(\mathcal C_h), h \in \mathcal I$, and $H^1_{de R}(\mathcal C_{\rm out})$ are trivial.
\end{lemma}
\noindent{\it Proof:} in  differential geometric terms Lemma \ref{lemm-a.8} means that every closed 1-differential form  in
$\mathcal C_h, h \in \mathcal I$, and  in $\mathcal C_{\rm out}$ is exact, what proves the lemma.

\begin{corollary}\label{cor-a.9}
The first groups of singular homology  $H_1(\mathcal C_h; \ere), h \in \mathcal I $  and  $H_1(\mathcal C_{\rm out}; \ere)$ are trivial.
\end{corollary}
\noindent{\it Proof:} The corollary follows from Lemma \ref{lemm-a.1} and Rham's theorem (Theorem 4.17 page 154 of \cite{w}).

\noindent{\bf Acknowledgement}

\noindent This research was partially done while M. Ballesteros was at  Departamento de M\'etodos Matem\'aticos  y Num\'ericos.
 Instituto de Investigaciones en Matem\'aticas Aplicadas y en Sistemas. Universidad Nacional Aut\'onoma de M\'exico.

\newpage

\begin{figure}
\begin{center}
\includegraphics[width=17cm]{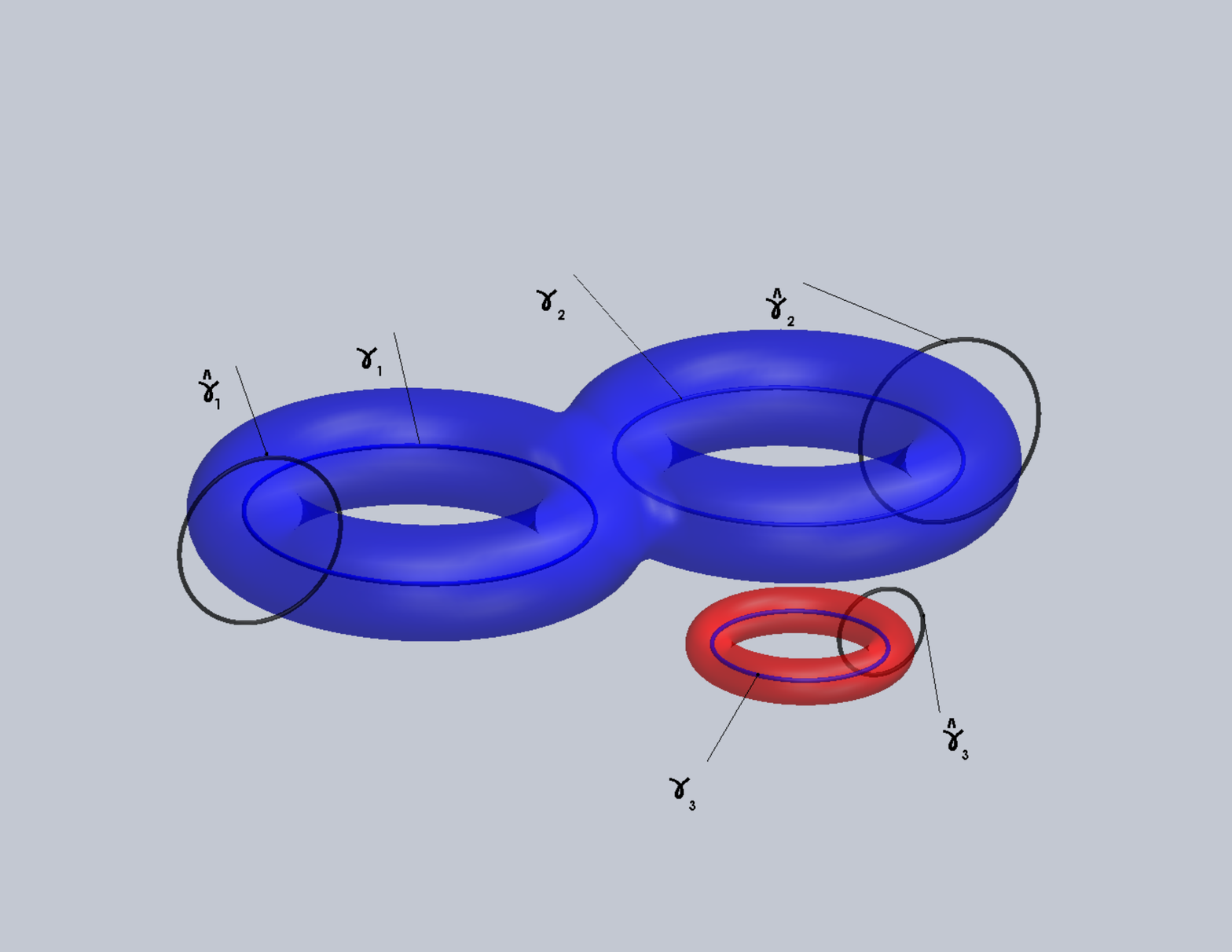}
\caption{The magnet  $ K= \cup_{j=1}^L  K_j \subset \ere^3$  where
$K_j$ are handlebodies , for
every $j \in \{ 1, \cdots, L \}$. The exterior domain,  $\Lambda:= \ere^3
\setminus K$.The curves $\gamma_k, k=1,2,\cdots m$ are a basis of
the first singular homology group of $K$ and the curves
 $\hat{\gamma}_k, k=1,2,\cdots m$ are a basis of the first singular homology group of $\Lambda$. }
\end{center}
\end{figure}
\newpage
\begin{figure}
\begin{center}
\includegraphics[height=13cm,width=15cm]{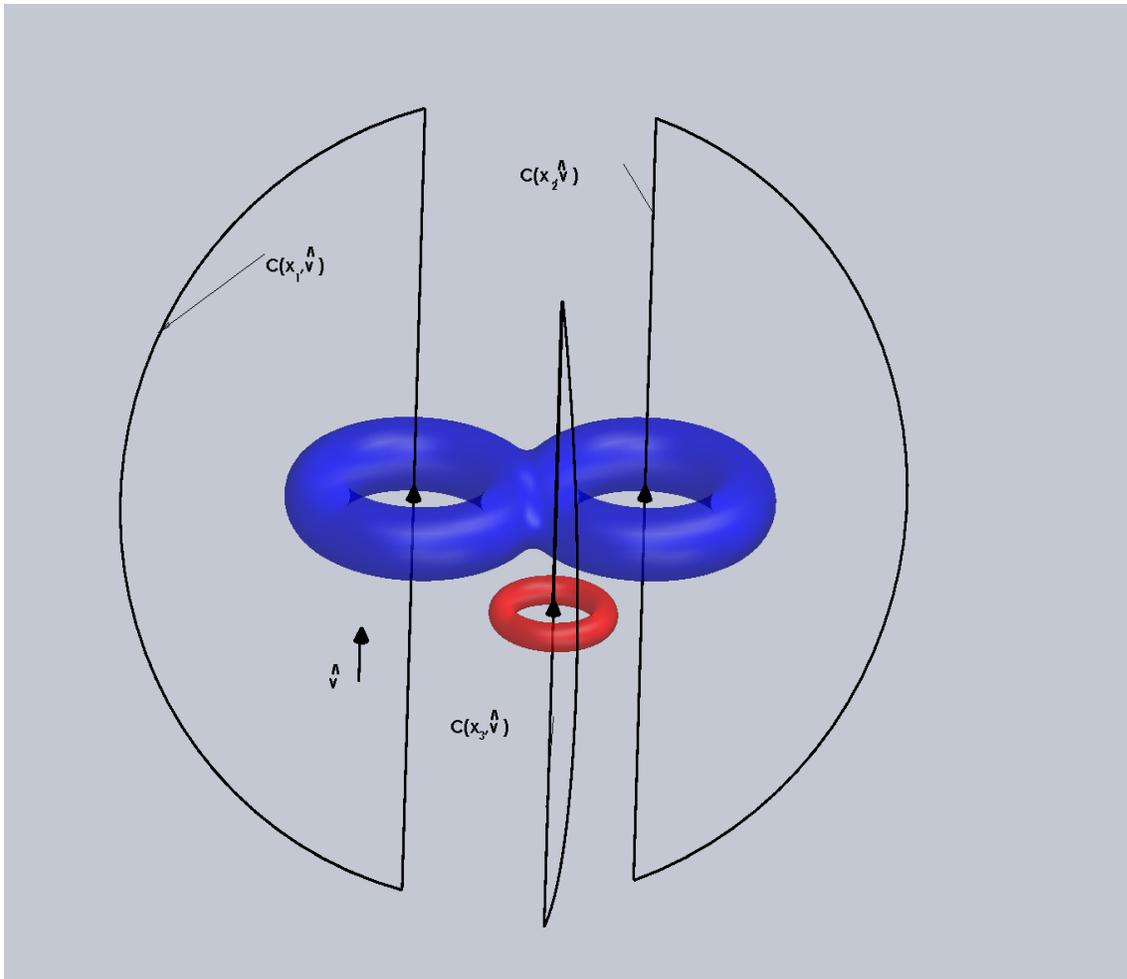}
\caption{The curves $c(x,\hv)$.}
\end{center}
\end{figure}

\end{document}